\documentclass{article}
\usepackage{arxiv}
\usepackage{hyperref}
\usepackage{latexsym}
\usepackage{url}
\usepackage[utf8]{inputenc}
\usepackage{booktabs}
\usepackage{amsmath}
\usepackage{amsfonts}
\usepackage{amssymb}
\usepackage[inline]{enumitem}
\usepackage{enumitem}
\usepackage{mathbbol}
\usepackage[T1]{fontenc}
\usepackage{mathtools}
\usepackage{multirow}
\usepackage{natbib}
\usepackage{rotating}
\usepackage{subcaption}
\usepackage{graphicx}
\usepackage{etaremune}
\usepackage{authblk}
\usepackage{tikz}

\newcommand{\ie}{\emph{i.e.}}
\newcommand{\eg}{\emph{e.g.}}

\newcommand{\vs}{\emph{vs. }}

\title{Overview of the TREC 2022 deep learning track}

\author[1]{Nick Craswell}
\author[1]{Bhaskar Mitra}
\author[2,3]{Emine Yilmaz}
\author[4,5]{Daniel Campos}
\author[6]{Jimmy Lin}
\author[7]{Ellen M. Voorhees}
\author[7]{Ian Soboroff}
\affil[1]{Microsoft\\\texttt{\small \{nickcr, bmitra\}@microsoft.com}}
\affil[2]{University College London\\\texttt{\small emine.yilmaz@ucl.ac.uk}}
\affil[3]{Amazon}
\affil[4]{University of Illinois Urbana-Champaign\\\texttt{\small dcampos3@illinois.edu}}
\affil[5]{Neural Magic Inc}
\affil[6]{University of Waterloo\\\texttt{\small jimmylin@uwaterloo.ca}}
\affil[7]{NIST\\\texttt{\small \{ellen.voorhees, ian.soboroff\}@nist.gov}}

\begin{document}
\maketitle

\begin{abstract}
This is the fourth year of the TREC Deep Learning track.
As in previous years, we leverage the MS MARCO datasets that made hundreds of thousands of human annotated training labels available for both passage and document ranking tasks.
In addition, this year we also leverage both the refreshed passage and document collections that were released last year leading to a nearly $16$ times increase in the size of the passage collection and nearly four times increase in the document collection size. Unlike previous years, in 2022 we mainly focused on constructing a more complete test collection for the passage retrieval task, which has been the primary focus of the track. The document ranking task was kept as a secondary task, where document-level labels were inferred from the passage-level labels.
Our analysis shows that similar to previous years, deep neural ranking models that employ large scale pretraining continued to outperform traditional retrieval methods.
Due to the focusing our judging resources on passage judging, we are more confident in the quality of this year's queries and judgments, with respect to our ability to distinguish between runs and reuse the dataset in future.
We also see some surprises in overall outcomes.
Some top-performing runs did not do dense retrieval. 
Runs that did single-stage dense retrieval were not as competitive this year as they were last year.
\end{abstract}
\section{Introduction}
\label{sec:intro}
At TREC 2022, we hosted the fourth TREC Deep Learning Track continuing our focus on benchmarking ad hoc retrieval methods in the large-data regime.
As in previous years~\citep{craswell2019overview, craswell2020overview, craswell2021overview}, we leverage the MS MARCO datasets~\citep{bajaj2016ms} that made hundreds of thousands of human annotated training labels available for both passage and document ranking tasks.
In addition, last year we refreshed both the passage and the document collections which also led to a nearly $16$ times increase in the size of the passage collection and nearly four times increase in the document collection size.
In addition to evaluating ranking methods on the larger collections, the data refresh also aimed at providing additional metadata---\eg, passage-to-document mappings---that may be useful for ranking, as well as incorporating some fixes for known text encoding issues in previous versions of the datasets.
This year we continue to benchmark against these larger passage and document collections.
However, the significant increase in collection sizes last year led to a corresponding increase in the number of relevant results in the collection per query and the existing judgment budget was exceeded before a reasonably complete set of these relevant results could be identified by the NIST judges.
This large number of relevant raised serious concerns about the test collection generated by last year's track, relating to reusability and also score saturation~\citep{voorhees2022too, craswell2021overview}.
To address these concerns, we made three changes this year with the goal of reducing the number of relevant results per query and in general the judgment costs so that they may be reused to obtain more complete set of judgments and consequently a more reusable test collection:
\begin{enumerate}
    \item We used test queries that did not contribute to the MS MARCO corpus. In all previous TREC DL and MS MARCO leaderboard evaluation, ten Bing results for the test query were included in the corpus whenever available. Further, we chose test queries where one of the Bing results was annotated as positive~\citep{gupta2022survivorship} and the positive result made it into our corpus. This year's queries went through the same MS MARCO sampling and top-10 annotation, but this happened after we finalized the MS MARCO dataset. We still choose queries that got a positive qrel during annotation, but the Bing top-10 passages and associated URLs were never used during corpus construction and we don't check whether the positive qrel is in the corpus. We no longer measure reciprocal rank, since that was the evaluation that used the MS MARCO qrels. Such qrels are still used for training and dev sets.
    \item We employ NIST judges to manually evaluate the relevance of retrieved results only for the passage ranking task and propagate the same labels to the source documents for the document ranking task.
    \item And finally, we detect near-duplicate passages and only judge one representative passage from each near-duplicate cluster with respect to the target query.
\end{enumerate}

This year we are more confident that our test collection is reusable and discriminative. We find results that confirm previous results, but also an overall larger gap between the best neural methods and traditional ranking methods. This could be due to the change in query sampling, but could also be due to progress in the field. We also see that there is a top run without dense retrieval, and the best run using single-stage dense retrieval is not as competitive as last year. Please see participant papers for more insights about what we learned this year.


\section{Task description}
\label{sec:task}

Similar to previous years, the Deep Learning Track in 2022 has two tasks: Passage ranking and document ranking. Participants were allowed to submit up to three official runs, and up to five additional baseline runs, for each task. When submitting each run, participants indicated what external data, pretrained models and other resources were used, as well as information on what style of model was used.

The TREC Deep Learning Track has a focus on generating reusable test collections and analyzing reusability. Since previous analysis showed that test collections constructed as part of the track in 2021 were not as reusable as the collections from the previous years~\citep{voorhees2022too, craswell2021trec}, in 2022 we primarily focused on improving the reusability of the test collections constructed as part of the track. Hence, we focused on the passage ranking task as the primary task (while keeping the document ranking task as the secondary task) and mainly aimed at constructing a sufficiently complete and reusable test collection for the passage ranking task. Labels inferred from passage-level labels have then been used for the document ranking task.

We changed our method for query sampling in 2022 with the intention of making the queries more difficult, to avoid the case where all runs have equally high performance and the evaluation is less discriminative. Since there was a risk the new queries would be unusable, we sampled $250$ backup queries using the same method as in 2021, and $250$ queries from a new method. Queries from the new method have query IDs of two million and above. Participants ran all $500$ queries. Our hope was that NIST judges would not find any problems with the new method, and could judge entirely queries from that set of $250$, and this was indeed the case.

Our new method uses queries from the same sampling and annotation pipeline as standard MS MARCO queries. The pipeline samples Bing queries, uses a classifier to find queries that are answerable by a short passage, and since the classifier is imperfect the annotators can also reject a query as ``can't judge''. For consistency with previous years, we also eliminated queries where the judge did not select a passage, see Figure\ref{fig:ms_marco_hit}. The difference is that all our MS MARCO ranking datasets until now were based on a 2018 version of the MS MARCO data with one million queries as described in a 2018 update of the MS MARCO paper \citep{bajaj2016ms}.\footnote{We note that the 2018 update of the paper \citep{bajaj2016ms} has an expanded author list, reflecting the expansion of the dataset to one million queries, which was planned by the original 2016 authors, and the addition of a ranking task, which was a new idea in 2018 not planned by the 2016 authors. The 2016 version and author list \citep{nguyen2016msmarco} reflect a preliminary release of the MS MARCO data, with 100 thousand queries and a natural language generation task.} This year's annotations went through the same process, but after the one million query cutoff. This means they were not in the one million MS MARCO queries, their top-10 passages and URLs were not used to construct the MS MARCO passage and document corpora. It also means we do not have an evaluation using the MS MARCO sparse qrels and we did not filter out test queries where the sparse qrel failed to make it into the corpus. We expect queries from the new method to be more difficult because 

In the pooling and judging process, NIST chose a subset of the $250$ queries for judging as described below. This led to a judged test set of $76$ queries for the passage ranking task, and we evaluated the document ranking runs on the same set of $76$ queries by propagating the passage labels to their source documents. 

Below we provide more detailed information about the document retrieval and passage retrieval tasks, as well as the datasets provided as part of these tasks.

\subsection{Passage ranking task}
The first task focuses on passage ranking, with two subtasks:
\begin{enumerate*}[label=(\roman*)]
    \item a full ranking and
    \item a top-$100$ reranking tasks.
\end{enumerate*}

In the full ranking subtask, given a query, the participants were expected to retrieve a ranked list of passages from the full collection based on the estimated likelihood of the passage containing an answer to the question.
Participants could submit up to $100$ passages per query for this end-to-end ranking task.

In the top-$100$ reranking subtask, $100$ passages per query were provided to participants, which were retrieved using Pyserini~\citep{lin2021pyserini}.
The reranking subtask allows all participants to start from the same starting point and to focus on learning an effective relevance estimator, without the need for implementing an end-to-end retrieval system.
It also makes the reranking runs more comparable, because they all rerank the same set of $100$ candidates.


This year's focus on building a more reusable test collection for the passage ranking task than the TREC~2021 collection caused changes in the assessment process at NIST.
One of the biggest changes was that only passages were judged, with passage judgments subsequently propagated to documents to form the document relevance judgments.
In previous years of the track, both documents and passages were judged independently, so focusing assessing resources on only passages effectively doubled the passage judgment budget.

The other major change was judging only a single element from a set of near-duplicate passages.
To effect this change, the passage corpus was clustered into classes of near-duplicate documents using the process at \url{https://github.com/isoboroff/dedupe}.
Each class had a single passage designated as the canonical passage for the class and the passage id of that passage was used as the class identifier.
The relevance label of the canonical passage with respect to a query was propagated to all the other passages in the same class.

The track received 100 submissions to the passage ranking task, 40 of which were baseline runs.
Of the 100 submitted runs, 82 runs contributed to the initial judgment pools.
The pool runs included all baseline runs, the three highest-priority submissions per team for the reranking subtask, and the three highest-priority submissions per team for the full ranking subtask.

The test set consisted of 500 queries, 250 of which (those whose query id is greater than 2,000,000) were queries that have no MSMARCO judgments.
NIST used this set of 250 new queries as candidates for judging.
Twenty-one of the candidates were eliminated by NIST staff before any judging took place because it was deemed unlikely to serve as a good evaluation query (e.g., {\em what a pull} and {\em what is my network name and password}); the remaining 229 candidates formed the set of queries that assessors could choose to work on.

An assessor chose a candidate query from the set and judged the first 100 passages (ordered by smallest rank at which the passage was retrieved across all pooled runs) in the depth-10 pool, or the entire depth-10 pool if the pool was smaller than 100 passages.
The candidate was discarded if at least 50\% of the judged passages were relevant or if no passage was relevant.
Otherwise, the assessor judged the remainder of the depth-10 pool (if any) plus the depth-10 pool formed on a fraction of the collection.

To support an (eventual) investigation of using corpus subsets in test collections, we needed to obtain judgments for the passages that would arise in such a case while the assessors were present.
This need motivated the use of both the full and subset corpora in the track judgment process.
The passages in the ``fractional'' collection were selected by randomly ordering the entire (deduped) passage corpus and using the first 1/10 of the passages in that ordering as the corpus.
The ordering was query independent, and the same ordering was used for all queries.
Runs were then restricted to the passages appearing in the 1/10 set by dropping any passage in the ranking but not in the fractional corpus from the ranking.
These are the {\em restricted} runs.
The passages added to the set of passages to be judged (the {\em judgment set}) were the depth-10 pools formed from the restricted runs in the pooled set, minus any passage already judged (because it was in the depth-10 pool of the full corpus).

The judgments from the first 100 passages were used to do an initial round of CAL processing~\citep{CAL}.
A CAL iteration takes all judgments made to date and ranks the remainder of the collection by likelihood of relevance.
While CAL was run across the corpus containing near duplicates, subsequent selection of passages to be added to the judgment set removed near duplicates, so only the canonical passage could be judged.
The first 25 passages in the deduped CAL ranking were also added to the judgment set.

A candidate query then went through a series of CAL iterations until a stopping condition was met.
If the relevant density (that is, the proportion of relevant passages to judged passages) was less than 40\%, at least 150 passages had been judged, and more than 3 relevant passages had been found, the candidate was accepted as a topic in the evaluation set.
If more than 300 passages had been judged and the relevant density was greater than 50\%, the candidate was stopped and rejected.
Otherwise, a candidate continued until the assessing budget had been expended.
Once the budget was spent, candidates with fewer than 150 passages judged, with fewer than four relevant passages found, or with a relevant density of at least 40\% were rejected.

Because CAL results depend on the set of judged passages given to it, we had three separate threads of CAL iterations running in parallel for each query that made it to the CAL stage.
One thread ran CAL using only passages in the fractional collection; the second thread ran CAL on the full corpus, but the judgments given to CAL contained only passages encountered in this thread; and the third thread used the full corpus and any available judgment.
The output of the third thread is the official qrels for the track.
Each CAL iteration added the next 25 passages from the ranking produced by CAL to the judgment set, except later iterations for the fractional thread which added 50 passages. 
The final evaluation set of topics contains the 76 topics accepted through this process.

Judgments were collected on a four-point scale:
\begin{etaremune}[start=3]
    \item \textbf{Perfectly relevant:} The passage is dedicated to the query and contains the exact answer.
    \item \textbf{Highly relevant:} The passage has some answer for the query, but the answer may be a bit unclear, or hidden amongst extraneous information.
    \item \textbf{Related:} The passage seems related to the query but does not answer it.
    \item \textbf{Irrelevant:} The passage has nothing to do with the query.
\end{etaremune}
For metrics that binarize the judgment scale, we map passage judgment levels 3,2 to relevant and map passage judgment levels 1,0 to irrelevant.

\subsection{Document ranking task}

Similar to the passage ranking task, the document ranking task focuses on two subtasks:
\begin{enumerate*}[label=(\roman*)]
    \item Full ranking and
    \item top-$100$ reranking.
\end{enumerate*}

The full ranking subtask models the end-to-end retrieval scenario, documents can be retrieved from the full document collection provided and the runs are expected to rank documents based on their relevance to the query. 

Similar to passage ranking, in the document reranking subtask, participants were provided with an initial ranking of $100$ documents, giving all participants the same starting point. The $100$ documents provided to the participants were generated using Pyserini~\cite{lin2021pyserini}. Participants were expected to rerank the 100 documents based on their estimated likelihood of containing an answer to the query.

Instead of collecting additional judgments for the document ranking task, we used passage judgments to infer judgments for documents: For each document we first identified the passages that were judged from within that document when collecting judgments for the passage ranking task, where all duplicates of a judged passage are assumed to have the same relevance judgment as the judged passage. If a document contains multiple passages with associated relevant judgments, we use the max judgment across all the passages to infer the final relevance judgment for the document. Previous work has shown that such an approach results in reasonable quality relevance judgments~\citep{Zhijing2019passage2doc}, and our study on the 2021 test collections further validated this~\citep{craswell2021overview}. 

Different from the passage ranking task, for document ranking metrics that use binary judgments we map document judgment levels 3,2,1 to relevant and map document judgment level 0 to irrelevant.

\section{Datasets}
\label{sec:data}

This year we leveraged the MS MARCO v2 dataset, which was used in both tasks. To understand how the new dataset differs from the old, we will first describe the natural language generation data and v1 ranking data.

\begin{figure}
    \centering
    \includegraphics[width=0.75\linewidth]{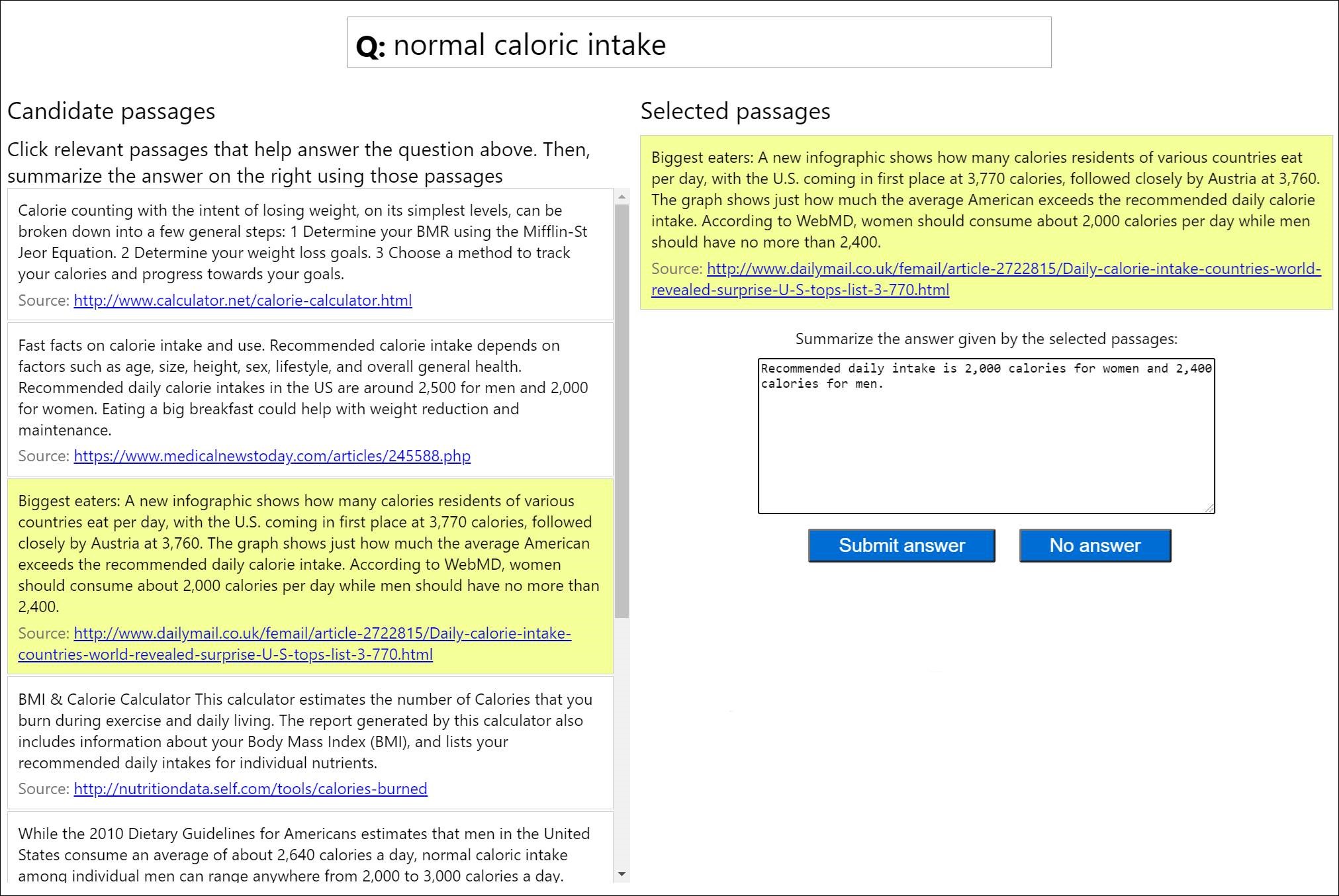}
    \caption{Crowd task used to generate the original MS MARCO natural language generation leaderboard. This same crowd data was later adapted to become the MS MARCO ranking tasks.}
    \label{fig:ms_marco_hit}
\end{figure}

\paragraph{MS MARCO natural language generation dataset.}
The original MS MARCO dataset was for a natural language generation task, rather than a ranking task. It processed one million queries, using a crowd task as shown in Figure~\ref{fig:ms_marco_hit}. The crowd worker would read the query, consider up to ten passages related to the query, decide if the passages could be used to answer the query and if answerable write an answer to the query in their own words. For each answerable question the crowd workers provided a non-extractive answer and an annotation of which passages they used to generate their answer. There was substantial quality work with the crowd workers to ensure quality and the crowd workers spent an average of 2.5 minutes on each annotation. The million queries were drawn from actual user queries to Bing. The ten results were generated by a Bing passage retrieval and ranking system. The queries were filtered before being annotated to remove any adult or offensive queries and any non-English queries. Moreover, further filtering was performed to ensure that the queries came from the $10-20\%$ of English queries that were detected as potentially being answerable with a short passage. Although the filter may be imperfect, the intention was to exclude navigational queries (such as [youtube]), queries that require a longer answer (such as [beef wellington recipe]) and queries that aim to complete some transaction (such as [buy xbox live]). We note that about 35\% of the queries could not be answered using the ten passages, in which case the crowd worker would indicate No answer, and one part of the original MS MARCO challenge was to predict which queries were answerable.

\paragraph{MS MARCO ranking v1 datasets.}
The MS MARCO passage and document ranking v1 datasets are used in the current MS MARCO leaderboards~\citep{lin2022fostering, craswell2021ms, lin2021significant} and in TREC 2019 and TREC 2020.

To generate the v1 passage ranking data, we took the union of the top ten passage lists for the one million queries, giving us 8.8 million distinct passages. For queries that were answerable, we used the crowd judge annotation for selected passages as a positive qrel. This gives us highly incomplete qrels, as noted in the original description \citep{bajaj2016ms}. We should in no way expect the positive qrel to be the ``best answer''. We found that training and evaluating using these sparse qrels gives us results that are quite correlated with results using much more comprehensive NIST judgments \citep{craswell2019overview, craswell2020overview}. Further study is needed to understand why this works, but we suspect it's important that the qrel is selected from a Bing ranking that has access to information that's unavailable to TREC participants, such as billions of past queries. This means the selected qrel is not biased towards some existing academic approach such as BM25. For each query that has a qrel, we generated a BM25 top-1000 for use in a reranking task and also allowed fullrank from the 8.8 million passages. We used the same split as in the QnA task: training ($80\%$), dev ($10\%$) and eval ($10\%$).

To generate the v1 document ranking data, we collected the corresponding urls for which the passages were extracted. Using these 3.5 million URLS, we obtained the associated document title and body corresponding to the ranking qrels. It is worth noting that the original passages were extracted between January 2016 and February 2018 while the full documents were extracted in March of 2018 and as a result only 3.2 million URLs were still able to be successfully extracted. From these documents, a \textit{clean} form was extracted where the body text had the HTML removed and focused on the main content of the page, removing web-page boilerplate such as navigation menus. Since we extracted the document text more than a year later than the passage data and used a completely different document parsing and processing pipeline (which unfortunately had character set processing issues) there was a chance that some pages that had a relevant passage no longer existed, no longer contained the passage, or even had the section of text with the passage accidentally removed as boilerplate. These are all realistic things to happen in a real-world application, where the document corpus is constantly changing, we do not wish to throw away our old relevance labels, and indeed we may not have budget to generate new labels. Doing a better job of generating a clean dataset using old labels is what we have now done in generating the v2 data. Qrels for the document task were assigned by assuming that a relevant passage qrel transfers to the document level as a positive document qrel. We generated top-100 document rankings using Indri, for use in a reranking task and also allowed fullrank from the 3.2 million documents.

The v1 data had several problems. The corpus was generated based on the queries, such that each passage and each document is in the corpus due to one of our million original queries. For each document in the corpus there may only be one passage in the passage dataset (and on average 2.8 passages per document), but that passage was identified by Bing in relation to one of the MS MARCO queries, possibly a test query. This is unrealistic, since a real system would be able to generate many candidate passages per document, and would not know what the test queries will be ahead of time. Therefore, we had to forbid participants from considering the passage-document mapping. The document dataset had several problems with character sets and missing whitespace.

\paragraph{MS MARCO ranking v2 datasets.}
The MS MARCO passage and document ranking v2 datasets were used for the first time in TREC 2021. The goal of the v2 dataset was to increase the scale and introduce a wider variety of documents such that not all documents were relevant to at least some query.

While the v1 data started with passages and was expanded to documents, the v2 data is document native. It begun by identifying documents based on the source urls of the v1 dataset. Of the original 3.5 million MS MARCO URLs, we were able to still find content for 2.7 million. We added an additional 9.2 million documents, selected to be the kind of documents that had useful passages of text in past Bing queries, giving a total of 11.9 million documents. For each document we ran a query-independent proprietary algorithm for identifying promising passages, and selected the best non-overlapping passages, giving on average 11.6 passages per document. This gives us our 138 million passages in the v2 passage corpus. We mapped the document qrels at the URL level, for training, dev and eval. The chance that the document is no longer relevant to the query, which also was a concern in v1 data, is now increased since the document content was extracted at a later date. We can consider how big this problem is by analyzing the disagreement rate between MS MARCO qrels and NIST qrels (in v1 and v2), and seeing whether training on MS MARCO qrels yields improved NIST NDCG on the test set. For mapping passage qrels, we required that the passage comes from the same URL as the original passage, and has sufficient text similarity to the positive passage text from v1. 

It is now possible for participants to use the passage-document mapping in participation, for example by considering document information in passage ranking, passage information in document ranking, and so on. Using a larger corpus prevents participants from proposing completely unscalable ranking approaches. The new dataset has fewer character encoding and whitespace issues, and could form the basis for future tasks that include some elements of additional document processing, such as extracting even shorter (phrase) answers.

\section{Results and analysis}
\label{sec:result}

\paragraph{Submitted runs}
A total of $14$ groups participated in the TREC 2022 Deep Learning Track.
Among them, $5$ groups participated in both the passage and the document ranking tasks, and of the remaining seven groups participated only in the passage ranking task and another two groups only in the document ranking task.
Similar to previous years, we also solicited baseline runs from the participating groups to enrich the judgment pools.
Across all groups, we received a total of $142$ run submissions, including $100$ passage ranking runs and $42$ document ranking runs.
This includes $59$ baseline runs---$40$ for passage ranking and $19$ for document ranking.
Table~\ref{tbl:runs-by-type} summarizes the submissions statistics for this year's track.

\begin{table}
    \centering
    \caption{TREC 2022 Deep Learning Track run submission statistics.}
    \begin{tabular}{lrr}
    \hline
    \hline
        & \textbf{Passage ranking} & \textbf{Document ranking} \\
        \hline
        Number of groups & 12 & 7 \\
        Number of total runs & 100 & 42 \\
        Number of baseline runs & 40 & 19 \\
        Number of runs w/ category: nnlm & 88 & 33 \\
        Number of runs w/ category: nn & 0 & 0 \\
        Number of runs w/ category: trad & 12 & 9 \\
        Number of runs w/ category: rerank & 11 & 3 \\
        Number of runs w/ category: fullrank & 89 & 39 \\
        Number of single-stage dense retrieval runs & 9 & 2 \\
        \hline
        \hline
    \end{tabular}
    \label{tbl:runs-by-type}
\end{table}

This year we had fewer participating groups ($14$ groups) compared to previous years ($15$ groups in 2019, $25$ in 2020, and $19$ in 2021).
However, we received a larger number of runs this year ($142$ runs) compared to previous years ($75$ runs in 2019, $123$ in 2020, and $129$ in 2021).
A larger number of baseline runs this year ($59$ runs) contributed towards this growth compared to previous years ($16$ runs in 2019, $34$ in 2020, and $37$ in 2021).
The number of official runs this year ($83$ runs) was slightly lower than the previous two years ($89$ runs in 2020 and $92$ in 2021) but higher than the inaugural year of the track ($59$ runs in 2019).

This year we asked participants to self-classify each of their runs under the following three categories (same taxonomy as was employed in our previous track overview papers~\citep{craswell2019overview, craswell2020overview, craswell2021overview}):
\begin{itemize}
    \item trad: No neural representation learning---\eg, classical learning to rank, PRF, and BM25
    \item nn: Representation learning with text as input, but not using a pre-trained model
    \item nnlm: Using a pre-trained model in any part of the pipeline---\eg, neural document expansion and BERT-style reranking
\end{itemize}

The largest category of runs was of type ``nnlm'' constituting $85\%$ of submissions across both tasks this year.
This was a significant increase over previous years---$44\%$ in 2019, $57\%$ in 2020, and $76\%$ in 2021---while the percentage of ``trad'' runs dipped this year to $15\%$ after having remained relatively stable over the previous years---$29\%$ in 2019, $33\%$ in 2020, and $24\%$ in 2021.
A significant shift also happened for the ``nn'' category over the previous years, decreasing from $27\%$ in 2019 to $10\%$ in 2020 and altogether disappearing as a category last year and this year.
This may reflect a convergence in the neural IR community, and the IR community in general, towards large language models, although whether this homogenization of approaches is healthy or premature is yet to be seen.

Participants were also asked to categorize their runs based on subtasks:
\begin{itemize}
    \item Rerank: Reranking the official top-100 candidates
    \item Fullrank: Full ranking from the collection (retrieval)
\end{itemize}

We observed an increase in the percentage of ``fullrank'' runs this year---$90\%$ compared to $72\%$ in 2019, $70\%$ in 2020, and $79\%$ in 2021.
The percentage of ``fullrank'' runs for the passage ranking task increased again this year---$89\%$ this year compared to $70\%$ in 2019, $69\%$ in 2020, and $81\%$ in 2021---which may have been partially influenced since last year by the reduction in size of the official reranking candidate set for the passage ranking task from $1000$ (as in the first two years of the track) to $100$ last year and this year.
The growing percentage of ``fullrank'' runs may also be due to increasing application of neural methods in the full ranking setting---either using dense retrieval methods~\citep{lee2019latent} or query term independent neural ranking models~\citep{mitra2019incorporating}.
The percentage of ``fullrank'' runs also increased for the document ranking task this year---$93\%$ this year compared to $74\%$ in 2019, $70\%$ in 2020, and $77\%$ in 2021.
Coincidentally, this year, we also asked participants to tell us
\begin{enumerate*}[label=(\roman*)]
    \item if their runs employed dense retrieval methods, and
    \item if the retrieval was performed in a single-stage under full retrieval setting.
\end{enumerate*}
We received $9$ single-stage dense retrieval runs for the passage ranking task this year and $2$ for the document ranking task.


\paragraph{Overall results}
Table~\ref{tab:passage_ranking_ranking} and Table~\ref{tab:document_ranking_ranking} present a standard set of relevance quality metrics for document and passage ranking runs, respectively, as we have reported for the track in previous years.
The reported metrics include Normalized Discounted Cumulative Gain (NDCG)~\citep{JK2002}, Normalized Cumulative Gain (NCG)~\citep{rosset2018optimizing}, and Average Precision (AP)~\citep{zhu2004recall}. These are all computed using NIST judgments, since this year's test queries do not have the sparse judgments that we used in previous years.

In subsequent discussions, we employ NDCG@10 as our primary evaluation metric to analyze ranking quality produced by different methods.
To analyze how different approaches compare beyond just the relevance of top-ranked results, we use NCG@100, which correlates more with how often relevant results are in the top-100 candidate set even if they are not eventually ranked as highly. 

\begin{table}[]
\caption{Summary of results for passage ranking runs. (For baselines see Appendix~\ref{sec:appendix}.)}
\scriptsize
\centering
\begin{tabular}{llllllrlr}
\toprule
run &      group &   subtask & neural &   stage & dense ret. &  NDCG@10 & NCG@100 &      AP \\
\midrule
pass3                   &        Ali &  fullrank &   nnlm &   multi &        yes &   0.7184 &  0.4313 &  0.2818 \\
NLE\_SPLADE\_CBERT\_DT5\_RR &        NLE &  fullrank &   nnlm &   multi &         no &   0.7145 &  0.4592 &  0.2950 \\
NLE\_SPLADE\_CBERT\_RR     &        NLE &  fullrank &   nnlm &   multi &         no &   0.7141 &  0.4565 &  0.2963 \\
pass2                   &        Ali &  fullrank &   nnlm &   multi &        yes &   0.7105 &  0.4007 &  0.2577 \\
NLE\_SPLADE\_RR           &        NLE &  fullrank &   nnlm &   multi &         no &   0.7092 &  0.4589 &  0.2977 \\
pass1                   &        Ali &  fullrank &   nnlm &   multi &        yes &   0.7050 &  0.4007 &  0.2442 \\
f\_sum\_mdt5              &     h2oloo &  fullrank &   nnlm &   multi &        yes &   0.7030 &  0.3993 &  0.2698 \\
srchvrs\_pz2\_colb2       &    srchvrs &  fullrank &   nnlm &   multi &        yes &   0.6630 &  0.3660 &  0.2160 \\
srchvrs\_ptn1\_colb2      &    srchvrs &  fullrank &   nnlm &   multi &        yes &   0.6562 &  0.3660 &  0.2066 \\
uogtr\_se\_gb             &      UoGTr &  fullrank &   nnlm &   multi &         no &   0.6508 &  0.3825 &  0.2252 \\
uogtr\_se\_gt             &      UoGTr &  fullrank &   nnlm &   multi &         no &   0.6508 &  0.3824 &  0.2256 \\
uogtr\_e\_gb              &      UoGTr &  fullrank &   nnlm &   multi &        yes &   0.6501 &  0.3818 &  0.2257 \\
uogtr\_be\_gb             &      UoGTr &  fullrank &   nnlm &   multi &         no &   0.6480 &  0.3558 &  0.2113 \\
srchvrs\_ptn2\_colb2      &    srchvrs &  fullrank &   nnlm &   multi &        yes &   0.6448 &  0.3660 &  0.2002 \\
srchvrs\_pz1\_colb2       &    srchvrs &  fullrank &   nnlm &   multi &         no &   0.6414 &  0.3501 &  0.2096 \\
srchvrs\_ptn1\_lcn\_colb2  &    srchvrs &  fullrank &   nnlm &   multi &         no &   0.6367 &  0.3501 &  0.1996 \\
uogtr\_e\_cprf\_t5         &      UoGTr &  fullrank &   nnlm &   multi &        yes &   0.6182 &  0.3621 &  0.2061 \\
yorku22a                &    yorku22 &  fullrank &   nnlm &   multi &        yes &   0.6089 &  0.3747 &  0.2003 \\
srchvrs\_p2\_colb2        &    srchvrs &  fullrank &   nnlm &   multi &        yes &   0.6010 &  0.3492 &  0.1745 \\
2systems                &        UGA &  fullrank &   nnlm &   multi &        yes &   0.5991 &  0.2958 &  0.1622 \\
unicoil\_reranked        &        UGA &  fullrank &   nnlm &   multi &        yes &   0.5910 &  0.2958 &  0.1605 \\
cip\_f2\_r                &        CIP &  fullrank &   nnlm &   multi &        yes &   0.5860 &  0.3393 &  0.1761 \\
cip\_f3\_r                &        CIP &  fullrank &   nnlm &   multi &        yes &   0.5852 &  0.3266 &  0.1708 \\
srchvrs\_p1\_colb2        &    srchvrs &  fullrank &   nnlm &   multi &         no &   0.5818 &  0.3400 &  0.1723 \\
srchvrs\_ptn3\_colb2      &    srchvrs &  fullrank &   nnlm &   multi &        yes &   0.5800 &  0.3660 &  0.1687 \\
6systems                &        UGA &  fullrank &   nnlm &   multi &        yes &   0.5783 &  0.3218 &  0.1604 \\
4systems                &        UGA &  fullrank &   nnlm &   multi &        yes &   0.5761 &  0.2959 &  0.1530 \\
c47                     &        UGA &  fullrank &   nnlm &   multi &        yes &   0.5701 &  0.2958 &  0.1493 \\
hierarchcal\_combination &        UGA &  fullrank &   nnlm &   multi &        yes &   0.5696 &  0.3554 &  0.1655 \\
uogtr\_s\_cprf            &      UoGTr &  fullrank &   nnlm &   multi &        yes &   0.5682 &  0.3501 &  0.1866 \\
p\_dhr                   &     h2oloo &  fullrank &   nnlm &  single &        yes &   0.5524 &  0.3420 &  0.1662 \\
graph\_colbert           &        UGA &  fullrank &   nnlm &   multi &        yes &   0.5482 &  0.3545 &  0.1656 \\
tuvienna-pas-col        &    DOSSIER &  fullrank &   nnlm &  single &        yes &   0.5386 &  0.3331 &  0.1677 \\
webis-dl-duot5-g        &      Webis &  fullrank &   nnlm &   multi &         no &   0.5314 &  0.1501 &  0.0887 \\
NLE\_ENSEMBLE\_SUM        &        NLE &    rerank &   nnlm &   multi &         no &   0.5286 &  0.1826 &  0.0948 \\
NLE\_ENSEMBLE\_CONDORCET  &        NLE &    rerank &   nnlm &   multi &         no &   0.5284 &  0.1826 &  0.0943 \\
p\_agg                   &     h2oloo &  fullrank &   nnlm &  single &        yes &   0.5282 &  0.3119 &  0.1461 \\
tuvienna-pas-unicol     &    DOSSIER &  fullrank &   nnlm &  single &        yes &   0.5231 &  0.3212 &  0.1518 \\
cip\_f1                  &        CIP &  fullrank &   nnlm &  single &        yes &   0.5121 &  0.3393 &  0.1469 \\
NLE\_T0pp                &        NLE &    rerank &   nnlm &   multi &         no &   0.5102 &  0.1826 &  0.0881 \\
fused\_3runs             &        UGA &    rerank &   nnlm &   multi &        yes &   0.5094 &  0.1826 &  0.0901 \\
uogtr\_t\_cprf            &      UoGTr &  fullrank &   nnlm &   multi &        yes &   0.5078 &  0.3250 &  0.1646 \\
yorku22b                &    yorku22 &  fullrank &   nnlm &  single &         no &   0.5076 &  0.2692 &  0.1130 \\
uogtr\_c\_cprf            &      UoGTr &  fullrank &   nnlm &   multi &        yes &   0.5075 &  0.2488 &  0.1355 \\
cip\_f1\_r                &        CIP &  fullrank &   nnlm &   multi &        yes &   0.5072 &  0.3563 &  0.1622 \\
fused\_2runs             &        UGA &    rerank &   nnlm &   multi &        yes &   0.5060 &  0.1826 &  0.0895 \\
hierarchical\_2runs      &        UGA &    rerank &   nnlm &   multi &        yes &   0.5001 &  0.1826 &  0.0885 \\
cip\_f2                  &        CIP &  fullrank &   nnlm &  single &        yes &   0.4997 &  0.3563 &  0.1429 \\
cip\_r2                  &        CIP &    rerank &   nnlm &   multi &         no &   0.4975 &  0.1826 &  0.0891 \\
webis-dl-duot5          &      Webis &  fullrank &   nnlm &   multi &         no &   0.4972 &  0.1501 &  0.0800 \\
webis-dl-duot5-aug-1    &      Webis &  fullrank &   nnlm &   multi &         no &   0.4925 &  0.1226 &  0.0781 \\
webis-dl-duot5-aug-2    &      Webis &  fullrank &   nnlm &   multi &         no &   0.4885 &  0.1226 &  0.0759 \\
Infosense-2             &  InfoSense &    rerank &   nnlm &   multi &         no &   0.4848 &  0.1826 &  0.0846 \\
cip\_f3                  &        CIP &  fullrank &   nnlm &  single &        yes &   0.4840 &  0.3266 &  0.1357 \\
Infosense-1             &  InfoSense &    rerank &   nnlm &   multi &         no &   0.4832 &  0.1826 &  0.0830 \\
cip\_r3                  &        CIP &    rerank &   nnlm &   multi &         no &   0.4669 &  0.1826 &  0.0795 \\
IELab-3MP-UT            &      ielab &  fullrank &   nnlm &  single &         no &   0.4658 &  0.2888 &  0.1101 \\
IELab-3MP-RBC           &      ielab &  fullrank &   nnlm &  single &         no &   0.4368 &  0.3220 &  0.1013 \\
cip\_r1                  &        CIP &    rerank &   nnlm &   multi &         no &   0.4320 &  0.1826 &  0.0719 \\
IELab-3MP-DI            &      ielab &  fullrank &   nnlm &  single &         no &   0.4148 &  0.2663 &  0.0832 \\
\bottomrule
\end{tabular}

\label{tab:passage_ranking}
\end{table}

\begin{table}[]
\caption{Summary of results for document ranking runs. (For baselines see Appendix~\ref{sec:appendix}.)}
\scriptsize
\centering
\begin{tabular}{llllllrlr}
\toprule
run &          group &   subtask & neural &   stage & dense ret. &  NDCG@10 & NCG@100 &      AP \\
\midrule
NLE\_SPLADE\_RR\_D           &            NLE &  fullrank &   nnlm &   multi &         no &   0.7611 &  0.5787 &  0.3453 \\
NLE\_SPLADE\_CBERT\_RR\_D     &            NLE &  fullrank &   nnlm &   multi &         no &   0.7601 &  0.5716 &  0.3387 \\
NLE\_SPLADE\_CBERT\_DT5\_RR\_D &            NLE &  fullrank &   nnlm &   multi &         no &   0.7598 &  0.5782 &  0.3405 \\
doc3                      &            Ali &  fullrank &   nnlm &   multi &        yes &   0.7488 &  0.5246 &  0.2997 \\
srchvrs\_dtn1              &        srchvrs &  fullrank &   nnlm &   multi &        yes &   0.5970 &  0.3492 &  0.1816 \\
NLE\_ENSEMBLE\_SUM\_doc      &            NLE &  fullrank &   nnlm &   multi &         no &   0.5918 &  0.2593 &  0.1619 \\
srchvrs\_dtn2              &        srchvrs &  fullrank &   nnlm &   multi &        yes &   0.5888 &  0.3492 &  0.1798 \\
NLE\_ENSEMBLE\_CONDORCE\_doc &            NLE &  fullrank &   nnlm &   multi &         no &   0.5882 &  0.2593 &  0.1609 \\
NLE\_T0pp\_doc              &            NLE &  fullrank &   nnlm &   multi &         no &   0.5843 &  0.2593 &  0.1587 \\
srchvrs\_d\_lb2             &        srchvrs &  fullrank &   nnlm &   multi &        yes &   0.5760 &  0.3492 &  0.1777 \\
srchvrs\_d\_lb1             &        srchvrs &  fullrank &   nnlm &   multi &        yes &   0.5754 &  0.3492 &  0.1782 \\
srchvrs\_d\_prd3            &        srchvrs &  fullrank &   nnlm &   multi &        yes &   0.5620 &  0.3492 &  0.1742 \\
srchvrs\_d\_prd1            &        srchvrs &  fullrank &   nnlm &   multi &        yes &   0.5546 &  0.3492 &  0.1705 \\
srchvrs\_d\_lb3             &        srchvrs &  fullrank &   nnlm &   multi &         no &   0.5302 &  0.2748 &  0.1407 \\
doc1                      &            Ali &  fullrank &   nnlm &   multi &        yes &   0.4936 &  0.4739 &  0.2154 \\
tuvienna                  &        DOSSIER &  fullrank &   nnlm &  single &        yes &   0.4868 &  0.3043 &  0.1294 \\
tuvienna-unicol           &        DOSSIER &  fullrank &   nnlm &  single &        yes &   0.4830 &  0.2985 &  0.1232 \\
doc2                      &            Ali &  fullrank &   nnlm &   multi &        yes &   0.4589 &  0.4739 &  0.2030 \\
ceqe\_custom\_rerank        &  CERTH\_ITI\_M4D &  fullrank &   nnlm &   multi &        yes &   0.3811 &  0.2599 &  0.1090 \\
rm3\_term\_filter\_rerank    &  CERTH\_ITI\_M4D &  fullrank &   nnlm &   multi &        yes &   0.3611 &  0.2425 &  0.1049 \\
plm\_128                   &     UAmsterdam &    rerank &   nnlm &   multi &         no &   0.3387 &  0.2236 &  0.0905 \\
plm\_64                    &     UAmsterdam &    rerank &   nnlm &   multi &         no &   0.3227 &  0.2236 &  0.0909 \\
plm\_512                   &     UAmsterdam &    rerank &   nnlm &   multi &         no &   0.2721 &  0.2236 &  0.0816 \\
\bottomrule
\end{tabular}

\label{tab:document_ranking}
\end{table}

\begin{figure}
  \center
  \begin{subfigure}{.49\textwidth}
    \includegraphics[width=\textwidth]{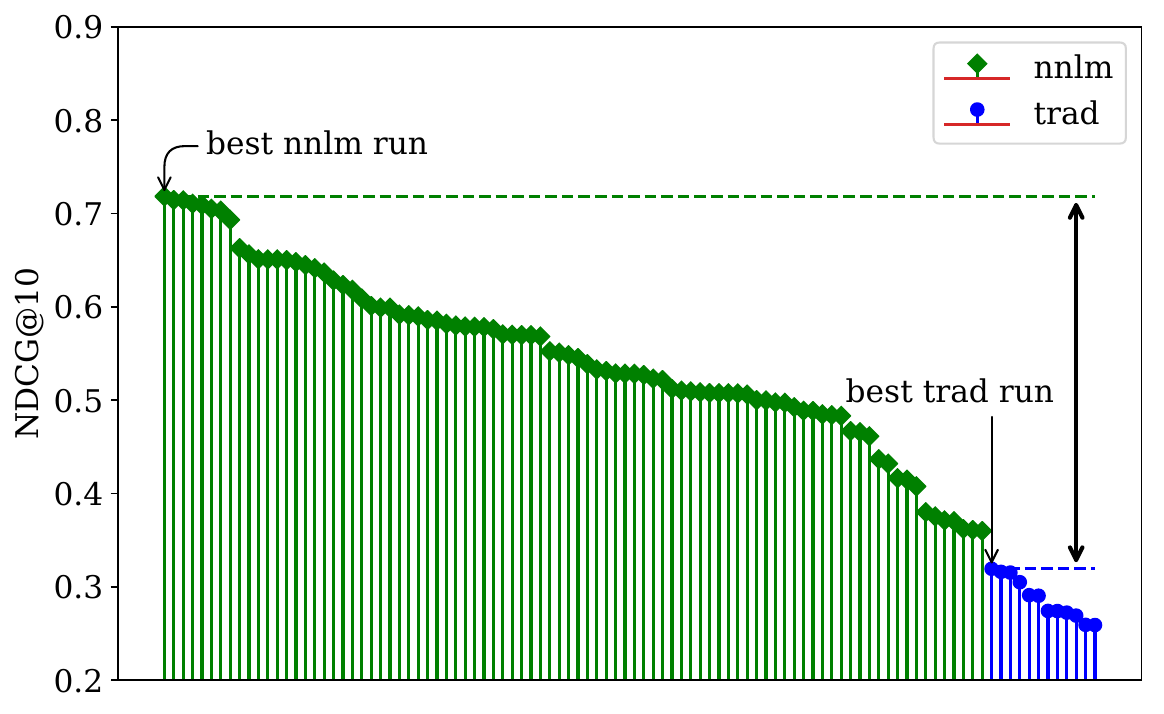}
    \caption{Passage ranking task}
    \label{fig:model-task-passages-stem-by-model-type}
  \end{subfigure}
  \hfill
  \begin{subfigure}{.49\textwidth}
    \includegraphics[width=\textwidth]{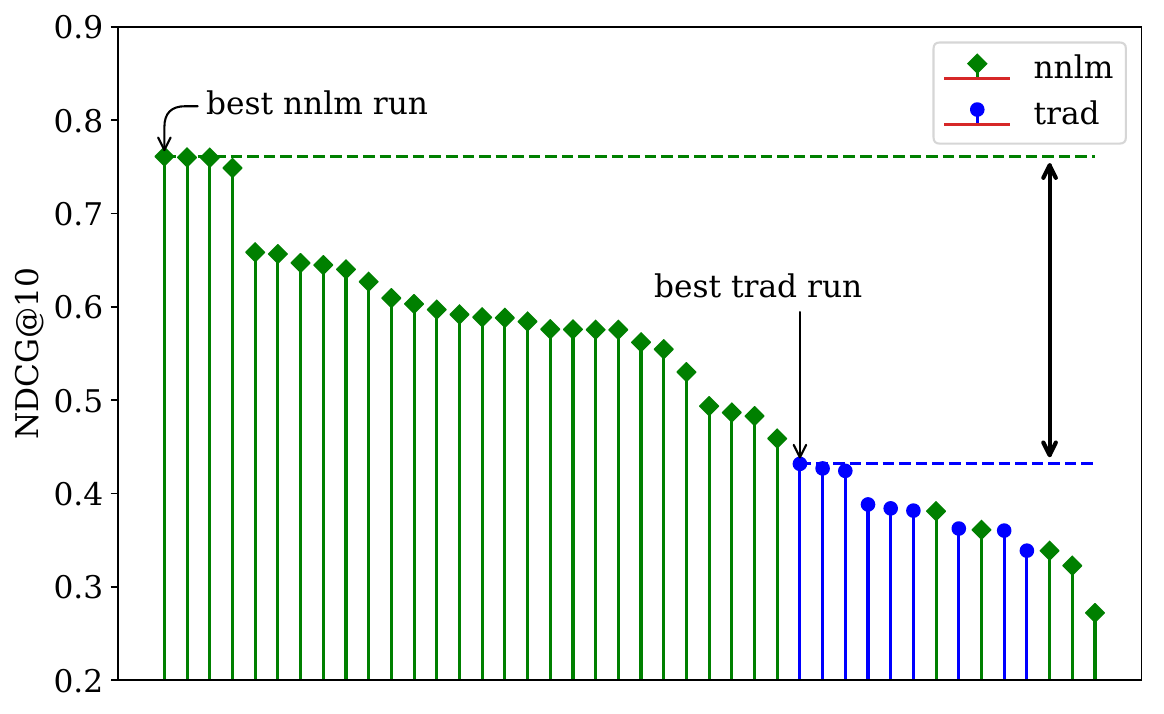}
    \caption{Document ranking task}
    \label{fig:model-task-docs-stem-by-model-type}
  \end{subfigure}
  \caption{NDCG@10 results by run type. As in the previous two years, ``nnlm'' runs continue to outperform over ``trad'' runs for both tasks.}
  \label{fig:model-stem-by-model-type}
\end{figure}

\paragraph{Neural \vs traditional methods.}
Figure~\ref{fig:model-stem-by-model-type} summarizes the evaluation results by run type---\ie, comparing ``nnlm'' \vs ``trad'' runs.
Across both document and passage ranking tasks, ``nnlm'' runs dramatically outperform ``trad'' runs this year.
For the passage ranking task, the best performing ``nnlm'' run improves NDCG@10 over the best performing ``trad'' run by $125\%$, while the same was $38\%$ in 2019, $42\%$ in 2020, and $36\%$ in 2021.
On the other hand, for the document ranking task, the NDCG@10 gap between the best performing run in `nnlm'' and ``trad'' categories is $76\%$ this year, compared to $29\%$ in 2019, $23\%$ in 2020, and $15\%$ in 2021.
Comparing percentage improvements across different year's tracks or across different tasks in the same year is not very meaningful due to differences in underlying data distributions.
However, we posit that the selection of more difficult test queries this year may have contributed to the seemingly increased gap between ``nnlm'' and ``trad'' run performances.

\begin{figure}
\includegraphics[width=\textwidth]{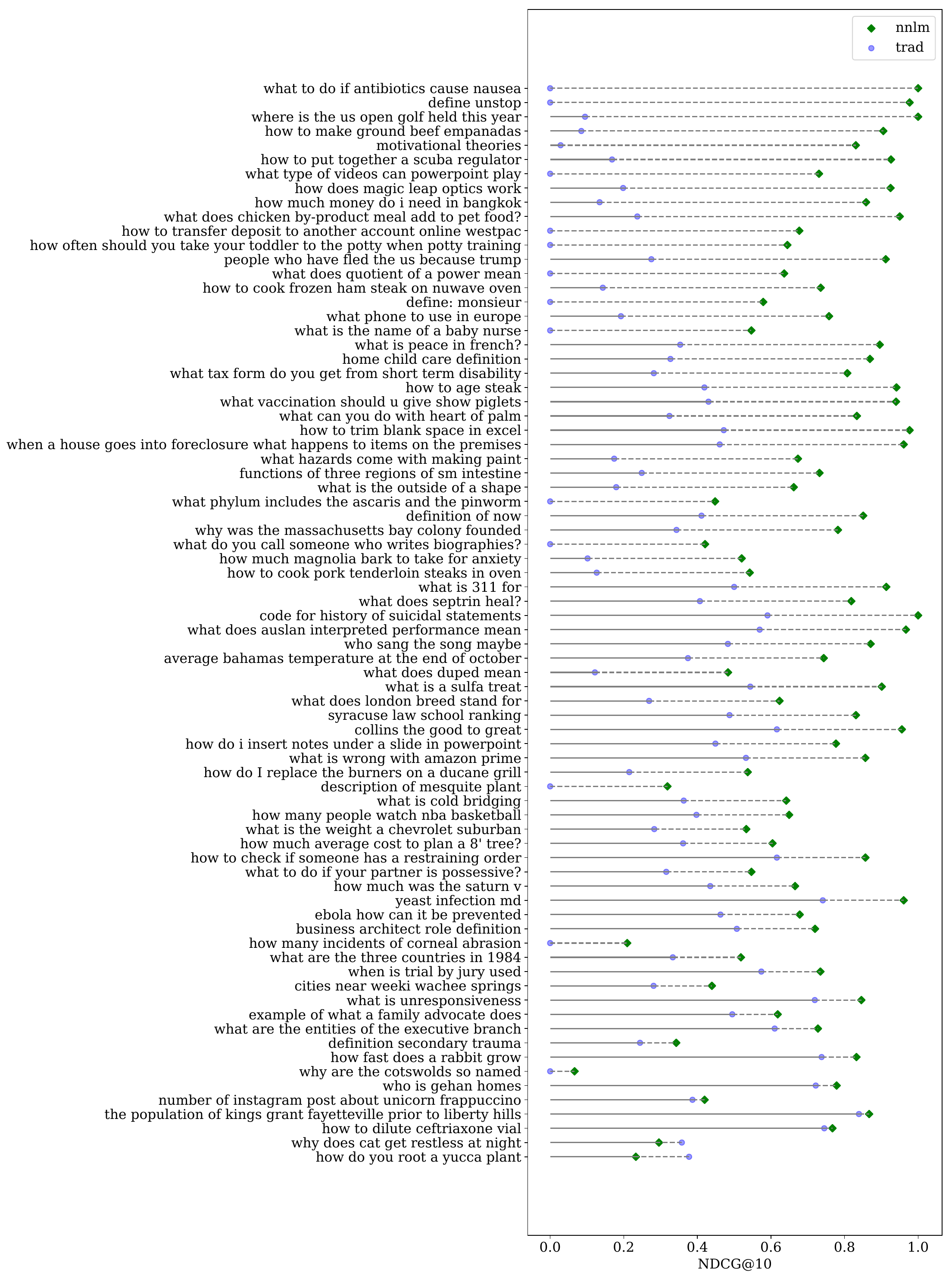}
\caption{Comparison of the best ``nnlm'' and ``trad'' runs on individual test queries for the passage ranking task. Queries are sorted by difference in mean performance between ``nnlm'' and ``trad'' runs. Queries on which ``nnlm'' wins with large margin are at the top.}
\label{fig:model-task-passages-bar-per-query}
\end{figure}

\begin{figure}
\includegraphics[width=\textwidth]{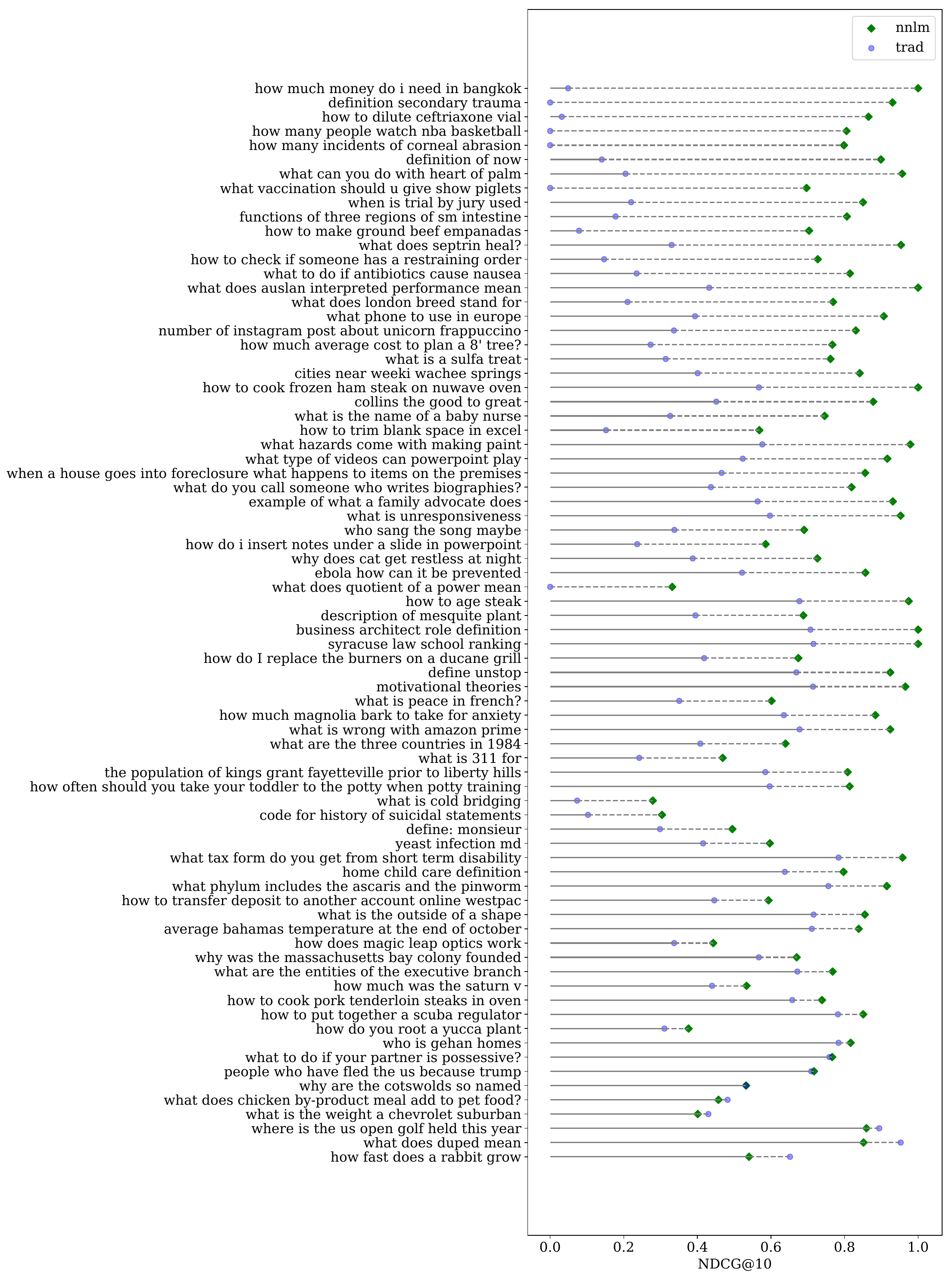}
\caption{Comparison of the best ``nnlm'' and ``trad'' runs on individual test queries for the document ranking task. Queries are sorted by difference in mean performance between ``nnlm'' and ``trad'' runs. Queries on which ``nnlm'' wins with large margin are at the top.}
\label{fig:model-task-docs-bar-per-query}
\end{figure}

Figure~\ref{fig:model-task-passages-bar-per-query} and \ref{fig:model-task-docs-bar-per-query} show a query-level comparison between the best ``nnlm'' and ``trad'' runs for the passage and the document ranking tasks, respectively.
The best ``nnlm'' run outperforms the best ``trad run'' on $74$ out of $76$ ($97\%$) queries for the passage ranking task---a big jump from $84\%$ in 2019, $88\%$ in 2020, and $89\%$ in 2021.
For the document ranking task, the best ``nnlm'' run wins on $71$ out of $76$ ($93\%$) queries against the best ``trad'' run, which is again much higher than $84\%$ in 2019 and 2020, and $72\%$ in 2021.

\begin{figure}
  \center
  \begin{subfigure}{.49\textwidth}
    \includegraphics[width=\textwidth]{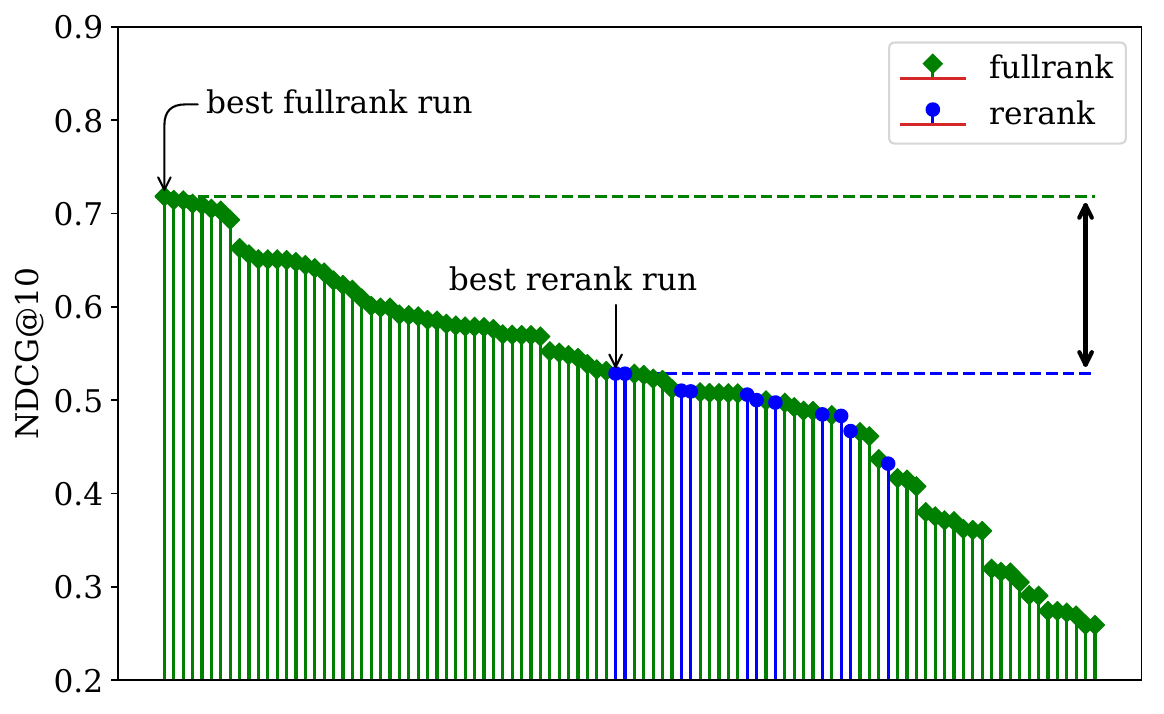}
    \caption{NDCG@10 for runs on the passage ranking task}
    \label{fig:model-task-passages-stem-by-subtask}
  \end{subfigure}
  \hfill
  \begin{subfigure}{.49\textwidth}
    \includegraphics[width=\textwidth]{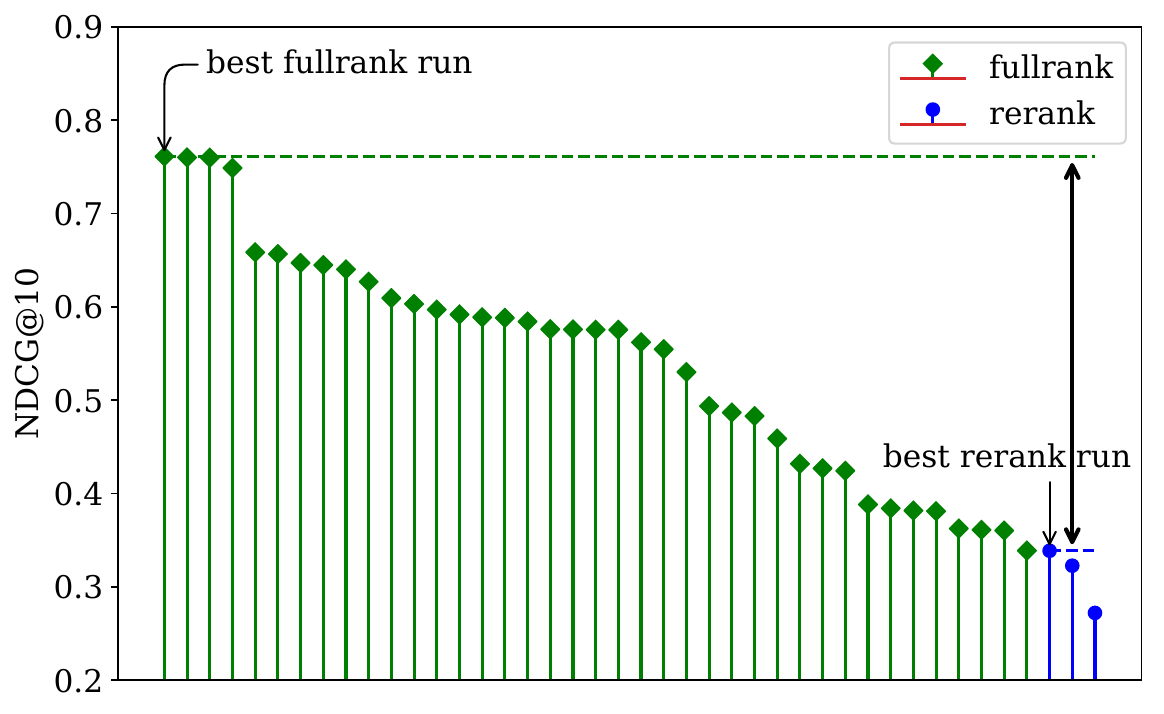}
    \caption{NDCG@10 for runs on the document ranking task}
    \label{fig:model-task-docs-stem-by-subtask}
  \end{subfigure}
  \begin{subfigure}{.49\textwidth}
    \includegraphics[width=\textwidth]{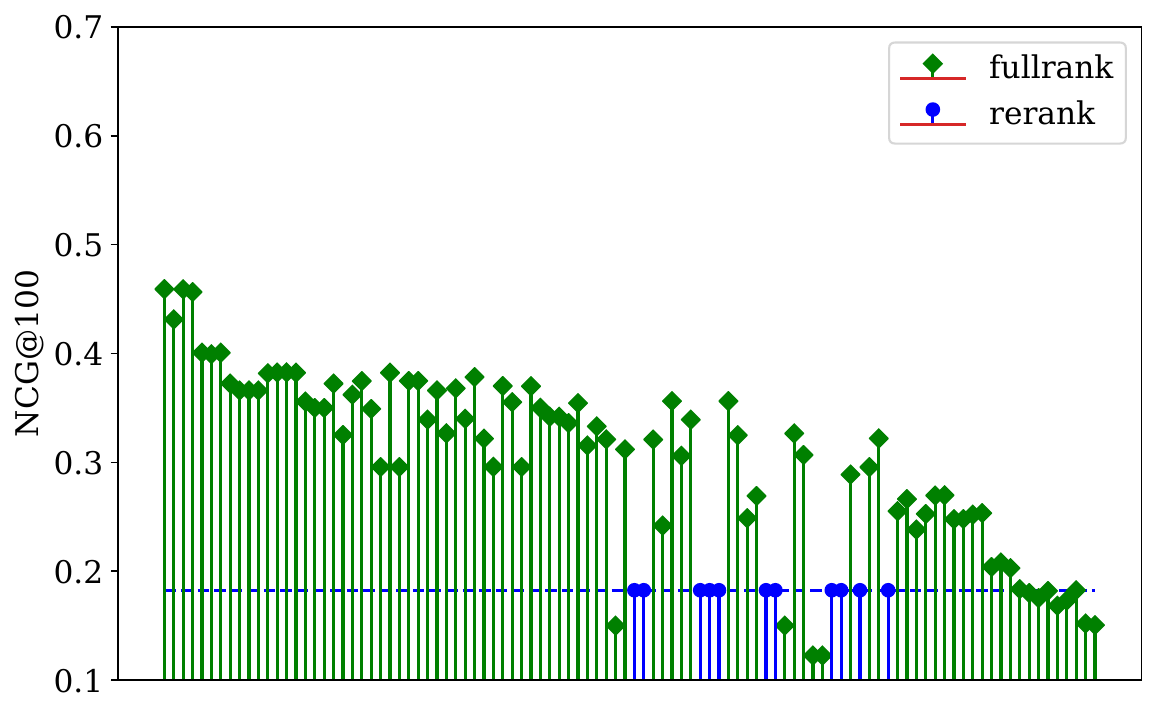}
    \caption{NCG@100 for runs on the passage ranking task}
    \label{fig:recall-task-passages-stem}
  \end{subfigure}
  \hfill
  \begin{subfigure}{.49\textwidth}
    \includegraphics[width=\textwidth]{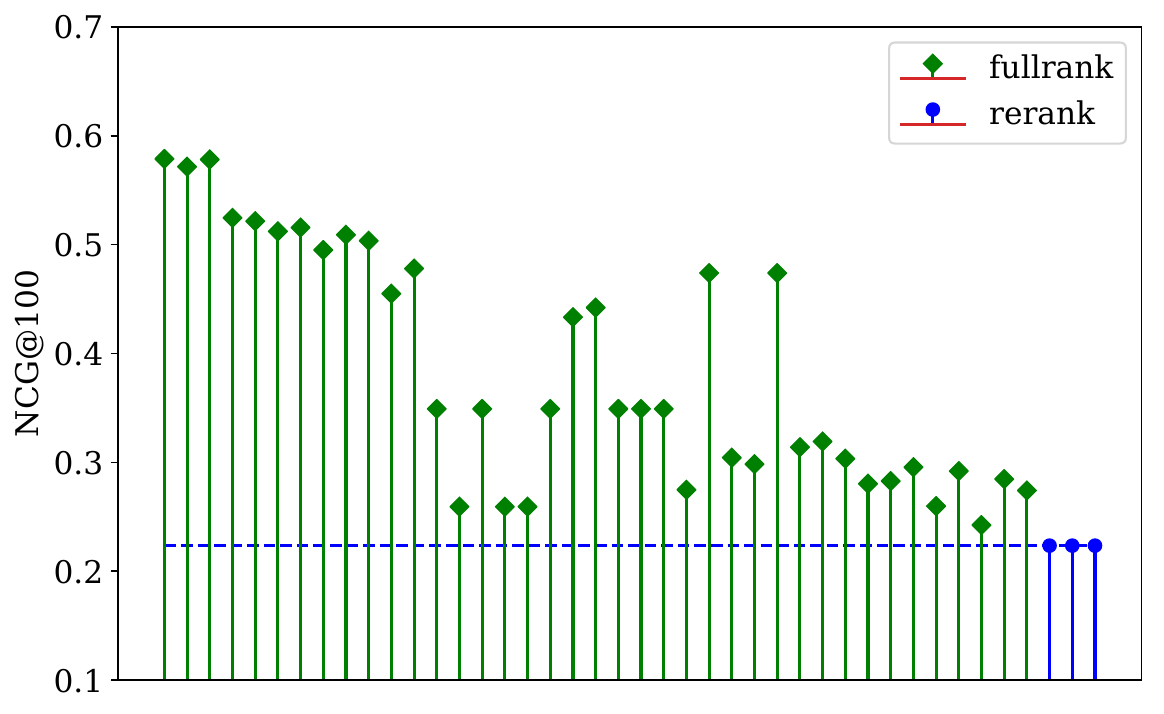}
    \caption{NCG@100 for runs on the document ranking task}
    \label{fig:recall-task-docs-stem}
  \end{subfigure}
  \caption{Comparing ``fullrank'' and ``rerank'' runs on ranking quality.
  Figure~(a) and (b) plots the NDCG@10 for different runs on the passage and document ranking tasks, respectively, and Figure~(c) and (d) plot the NCG@100 for the same.
  We order the runs by their NDCG@10 performance along the $x$-axis in all four plots.
  The best run for both tasks correspond to the ``fullrank'' setting.}
  \label{fig:recall-stem}
\end{figure}

\paragraph{Full ranking \vs reranking.}
This year for the passage ranking task, the best ``fullrank'' run has a $36\%$ NDCG@10 improvement over the best ``rerank'' run, compared to $4\%$ improvement in 2019, no improvement in 2020, and $6\%$ improvement in 2021.
For the document task this year, the best ``fullrank'' run has $125\%$ higher NDCG@10 than the best ``rerank'' run, which we can compare with a $1\%$ improvement in 2019, $5\%$ improvement in 2020, and $4\%$ improvement in 2021.
If we compare Figure~\ref{fig:recall-stem} (a) and (c) (and similarly Figure~\ref{fig:recall-stem} (b) and (d)), we also notice a stronger correlation between NDCG@10 and NCG@100 metrics compared to previous years.
While we reiterate that comparing percentage improvements across different tasks and across different years are not very meaningful, we note that these differences between best performing ``fullrank'' and ``rerank'' are particularly large this year compared to previous years of the track.
There may be many contributing factors including, but not limited to:
\begin{enumerate*}[label=(\roman*)]
    \item Potential recent progress by the community in the ``fullrank'' setting,
    \item increased difficulty of this year's test set, and/or
    \item less interest from participating groups this year in the ``rerank'' setting leading to under-optimized ``rerank'' runs.
\end{enumerate*}

This year the best single-stage dense retrieval run was $23\%$ worse on NDCG@10 compared to the best passage ranking run and $36\%$ worse on NDCG@10 compared to the best document ranking run.
Again these numbers are very different compared to last year's where the best single-stage dense retrieval run was behind the best run on NDCG@10 by only $10\%$ for passage ranking and $6\%$ for document ranking.



\paragraph{Effect of near duplicates}
The main motivation for processing near duplicates was to increase the number of distinct passages that would be seen in the assessing process.
During judging we used deduped runs, that had been processed to retain only a single instance of each near-dupe cluster.

To think about the effect of dupes, we can consider four approaches:
\begin{itemize}
    \item Do nothing: In previous years we left the near-dupes in the runs and potentially judged multiple results from the same near-dupe cluster.
    \item Dedupe corpus: It would be possible to dedupe the corpus ahead of time, reducing each near-dupe cluster to a single canonical passage ID, and perhaps somehow patching the document-passage mapping to still have a coherent collection. We have never done corpus-level deduping in any task for this track.
    \item Dedupe runs: During judging we mapped all passages from each near-dupe cluster to a single canonical passage ID. The evaluation with deduped runs should be similar to that for deduping the corpus, but not identical, since deduped top-100 lists may no longer have 100 results and some ranking methods may change when corpus statistics change.
    \item Expand qrels: After judging with dedupe runs, we expanded the qrels so that every passage in each labeled near-dupe cluster gets the same label, although judges only saw one of them. The evaluation with expanded qrels should be similar to the ``do nothing'' case above, since runs contain dupes and multiple near-dupe results can be labeled. Qrel expansion can generate a very large number of labels if something is labeled from a very large near-dupe cluster.
\end{itemize}
To examine the effect of duplicates, we compared evaluation with expanded qrels to evaluation with deduped runs. 


\begin{table}
    \footnotesize
    \centering
    \caption{Per-query counts of number of judged passages and respective labels for the deduped runs (O) and expanded qrels (E).}
    \begin{tabular}{|l|rr|rr|rr|rr||l|rr|rr|rr|rr|}
    \hline
    \multicolumn{1}{|c|}{Query} & \multicolumn{2}{c|}{Total} &  \multicolumn{2}{c|}{Level 1} & \multicolumn{2}{c|}{Level 2} & \multicolumn{2}{c||}{Level 3} & \multicolumn{1}{c|}{Query} & \multicolumn{2}{c|}{Total} &  \multicolumn{2}{c|}{Level 1} & \multicolumn{2}{c|}{Level 2} & \multicolumn{2}{c|}{Level 3} \\
     & \multicolumn{1}{c}{O} & \multicolumn{1}{c|}{E} & \multicolumn{1}{c}{O} & \multicolumn{1}{c|}{E} & \multicolumn{1}{c}{O} & \multicolumn{1}{c|}{E}  & \multicolumn{1}{c}{O} & \multicolumn{1}{c||}{E}  & & \multicolumn{1}{c}{O} & \multicolumn{1}{c|}{E} & \multicolumn{1}{c}{O} & \multicolumn{1}{c|}{E} & \multicolumn{1}{c}{O} & \multicolumn{1}{c|}{E}  & \multicolumn{1}{c}{O} & \multicolumn{1}{c|}{E}\\
     \hline
2000511 & 252 & 1492 & 62 & 149 & 17 & 17 & 0 & 0 & 2030608 & 309 & 453 & 25 & 32 & 16 & 23 & 17 & 21 \\
2000719 & 286 & 42710 & 115 & 127 & 41 & 43 & 47 & 52 & 2031726 & 425 & 43117 & 279 & 42865 & 107 & 180 & 0 & 0 \\
2001532 & 319 & 369 & 107 & 112 & 74 & 92 & 30 & 38 & 2032090 & 256 & 267 & 120 & 128 & 34 & 34 & 0 & 0 \\
2001908 & 226 & 276 & 9 & 9 & 17 & 18 & 0 & 0 & 2032949 & 384 & 460 & 73 & 82 & 39 & 44 & 47 & 57 \\
2001975 & 333 & 406 & 127 & 152 & 35 & 46 & 68 & 74 & 2032956 & 347 & 403 & 5 & 5 & 11 & 11 & 1 & 1 \\
2002146 & 224 & 270 & 13 & 15 & 40 & 49 & 0 & 0 & 2033232 & 361 & 387 & 8 & 10 & 10 & 11 & 1 & 1 \\
2002269 & 282 & 312 & 134 & 149 & 51 & 54 & 24 & 27 & 2033396 & 261 & 677 & 104 & 129 & 65 & 73 & 20 & 23 \\
2002533 & 423 & 888 & 2 & 2 & 9 & 10 & 0 & 0 & 2033470 & 321 & 583 & 66 & 69 & 42 & 44 & 83 & 104 \\
2002798 & 266 & 322 & 181 & 226 & 18 & 22 & 0 & 0 & 2034205 & 356 & 461 & 102 & 123 & 75 & 100 & 0 & 0 \\
2003157 & 222 & 246 & 138 & 153 & 2 & 2 & 8 & 11 & 2034676 & 339 & 1243 & 44 & 424 & 13 & 19 & 2 & 2 \\
2003322 & 232 & 369 & 92 & 174 & 25 & 34 & 55 & 75 & 2035009 & 381 & 410 & 26 & 29 & 8 & 10 & 5 & 6 \\
2003976 & 235 & 271 & 47 & 56 & 36 & 41 & 2 & 3 & 2035447 & 345 & 1421 & 4 & 5 & 23 & 25 & 1 & 1 \\
2004237 & 331 & 414 & 169 & 201 & 61 & 78 & 23 & 24 & 2035565 & 281 & 341 & 62 & 70 & 25 & 31 & 39 & 57 \\
2004253 & 268 & 283 & 149 & 158 & 54 & 58 & 19 & 20 & 2036182 & 318 & 4615 & 72 & 74 & 12 & 12 & 0 & 0 \\
2005810 & 317 & 513 & 61 & 90 & 57 & 82 & 7 & 11 & 2036968 & 248 & 328 & 96 & 128 & 53 & 69 & 27 & 35 \\
2005861 & 366 & 773 & 29 & 31 & 11 & 11 & 42 & 47 & 2037251 & 338 & 376 & 29 & 39 & 38 & 40 & 47 & 49 \\
2006211 & 241 & 296 & 100 & 116 & 10 & 12 & 8 & 10 & 2037609 & 399 & 42864 & 51 & 56 & 21 & 25 & 4 & 5 \\
2006375 & 252 & 42677 & 78 & 89 & 25 & 27 & 74 & 81 & 2037924 & 354 & 403 & 12 & 12 & 66 & 72 & 11 & 13 \\
2006394 & 340 & 430 & 60 & 77 & 34 & 51 & 0 & 0 & 2038466 & 266 & 315 & 233 & 282 & 11 & 11 & 9 & 9 \\
2006627 & 284 & 637 & 31 & 36 & 22 & 22 & 27 & 29 & 2038890 & 246 & 720 & 4 & 183 & 1 & 1 & 9 & 9 \\
2007055 & 304 & 444 & 137 & 203 & 76 & 98 & 0 & 0 & 2039908 & 324 & 417 & 59 & 68 & 17 & 23 & 77 & 99 \\
2007419 & 340 & 391 & 96 & 109 & 19 & 21 & 17 & 18 & 2040287 & 177 & 204 & 30 & 33 & 24 & 28 & 25 & 30 \\
2008871 & 382 & 460 & 128 & 153 & 120 & 141 & 2 & 2 & 2040352 & 255 & 378 & 1 & 1 & 4 & 6 & 24 & 46 \\
2009553 & 216 & 258 & 5 & 5 & 5 & 5 & 1 & 1 & 2040613 & 285 & 358 & 37 & 38 & 32 & 33 & 23 & 24 \\
2009871 & 282 & 341 & 106 & 126 & 35 & 43 & 55 & 59 & 2043895 & 318 & 350 & 171 & 185 & 95 & 107 & 11 & 13 \\
2012431 & 318 & 412 & 87 & 98 & 54 & 80 & 34 & 46 & 2044423 & 143 & 178 & 6 & 6 & 4 & 5 & 7 & 12 \\
2012536 & 416 & 491 & 245 & 285 & 108 & 133 & 0 & 0 & 2045272 & 423 & 487 & 94 & 114 & 109 & 121 & 5 & 5 \\
2013306 & 313 & 43946 & 26 & 31 & 4 & 42396 & 10 & 12 & 2046371 & 346 & 486 & 108 & 144 & 124 & 166 & 8 & 13 \\
2016333 & 320 & 351 & 74 & 78 & 19 & 22 & 1 & 1 & 2049417 & 366 & 471 & 39 & 48 & 24 & 30 & 5 & 8 \\
2017299 & 265 & 306 & 185 & 214 & 3 & 4 & 5 & 5 & 2049687 & 258 & 312 & 47 & 50 & 10 & 11 & 13 & 15 \\
2025747 & 336 & 49627 & 28 & 1776 & 23 & 24 & 25 & 33 & 2053884 & 255 & 276 & 3 & 3 & 27 & 28 & 0 & 0 \\
2026150 & 320 & 384 & 140 & 168 & 36 & 42 & 24 & 24 & 2054355 & 315 & 387 & 70 & 81 & 37 & 38 & 3 & 4 \\
2026558 & 355 & 42950 & 41 & 46 & 20 & 24 & 3 & 4 & 2055211 & 331 & 386 & 36 & 44 & 45 & 49 & 2 & 2 \\
2027130 & 332 & 459 & 100 & 125 & 41 & 53 & 80 & 92 & 2055480 & 240 & 329 & 14 & 18 & 6 & 8 & 6 & 6 \\
2027497 & 409 & 499 & 187 & 228 & 76 & 100 & 0 & 0 & 2055634 & 240 & 292 & 122 & 144 & 39 & 46 & 10 & 13 \\
2028378 & 381 & 486 & 111 & 132 & 88 & 129 & 48 & 64 & 2055795 & 350 & 387 & 143 & 153 & 92 & 95 & 20 & 20 \\
2029260 & 424 & 44035 & 112 & 118 & 102 & 114 & 47 & 49 & 2056158 & 476 & 535 & 104 & 121 & 159 & 179 & 2 & 2 \\
2030323 & 242 & 274 & 75 & 84 & 62 & 68 & 33 & 39 & 2056323 & 231 & 271 & 106 & 121 & 5 & 6 & 2 & 3 \\
    \hline
    \end{tabular}
    \label{qrels.tab}
\end{table}

The expanded qrels, which are the official qrels, have 386,416 judgments. Contributing to the large number of qrels are a few very large duplicate clusters. Table~\ref{qrels.tab} shows the total number of passages with judgments for deduped runs and expanded qrels, as well as the number of passages with a non-zero relevance value in each qrels.
For eight topics there is some judgment for a large duplicate class, leading to a large number of total expanded (E) qrels. One topic has a that class judged at level 1 and another topic has that class judged at level 2, but for most topics we either saw no large class or a large class at judgment level 0.

\begin{figure}
   \centering
   \includegraphics[width=0.45\linewidth]{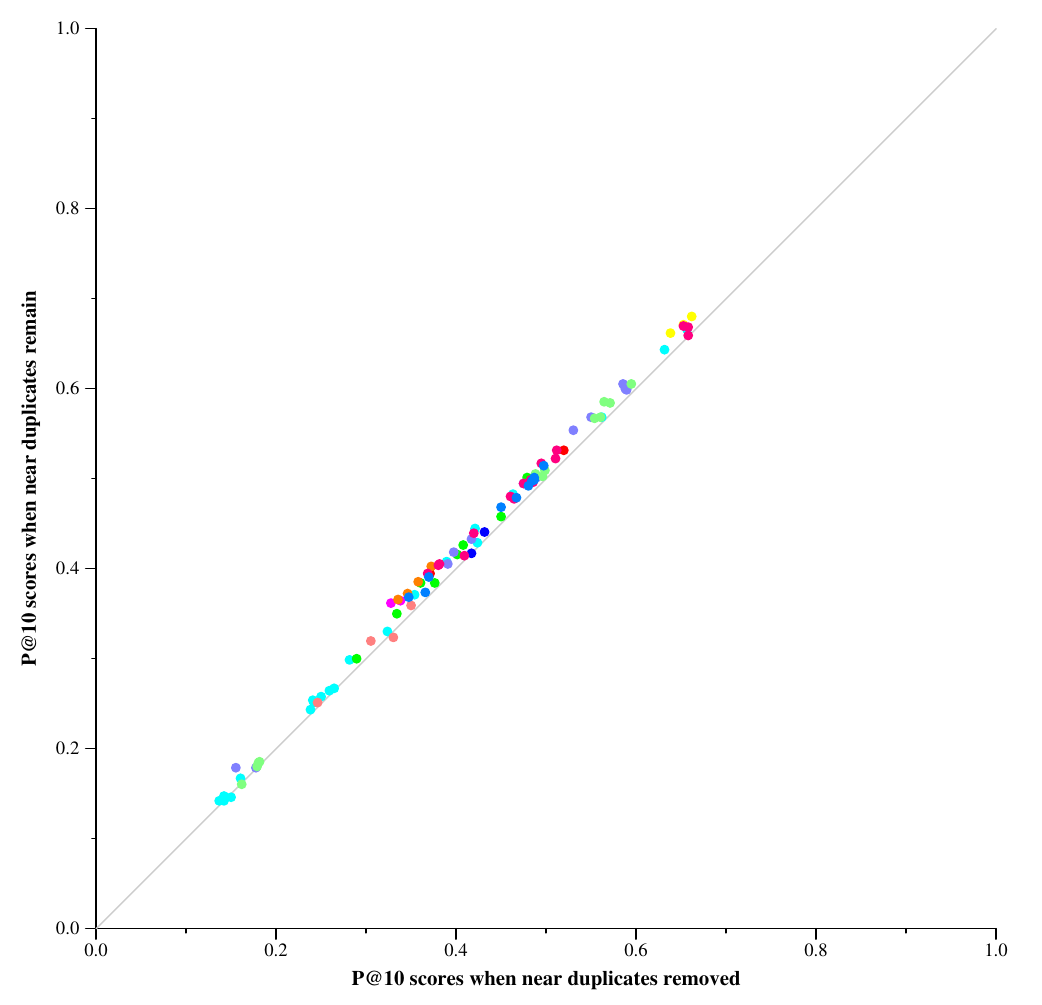}
   \includegraphics[width=0.45\linewidth]{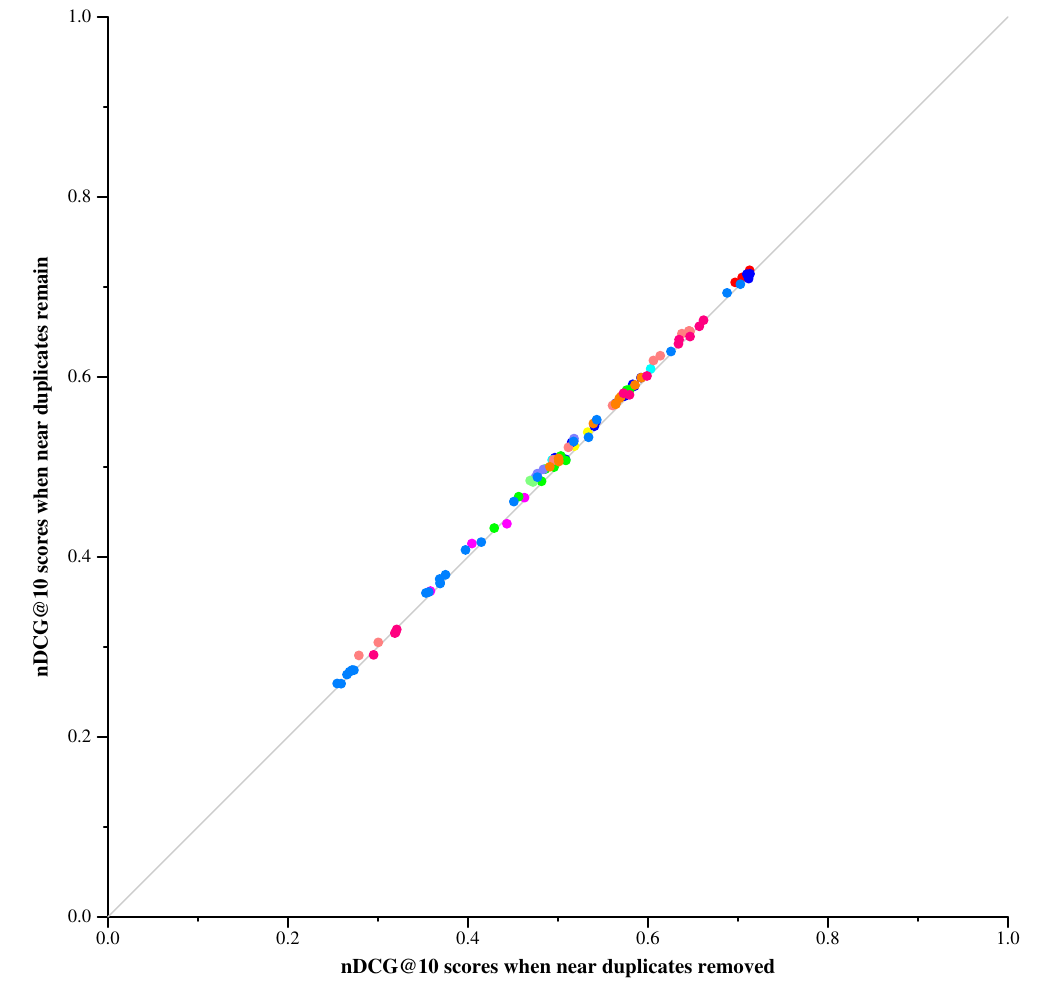}
   \caption{Changes in P@10 (left) and nDCG@10 (right) scores when passage ranking submissions are scored using or not using near duplicate passages. Each dot represents one run and runs with the same color dot were submitted by the same group.}
   \label{dupes.fig}
\end{figure}

The effect of deduped runs vs expanded qrels on system scoring is shown in Figure~\ref{dupes.fig}.
In the figure, the score computed for a run using deduped runs is plotted on the x-axis while the score computed for the run using the expanded qrels  is plotted on the y-axis.
Each dot in the plot represents one run, and each run submitted by a given group is plotted in the same color.
P@10 scores are shown in the graph on the left of the figure, and nDCG@10 scores are shown in the graph on the right.
The absolute value of the nDCG scores are very similar using the two qrels as evidenced by most dots lying on the diagonal line; relative scores between runs thus also preserved.
There are somewhat more differences in scores for P@10, though the differences are still not very large and relative scores are still mostly stable.
The fact that the absolute value of the scores is greater for the expanded qrels (most dots are above the line) for P@10 is an indication that runs did retrieve multiple instances of relevant near-duplicate passages in the top 10 ranks.

Overall, a major disadvantage of the deduped runs approach is that any future use of this year's test collection would require the runs also to be deduped, adding complexity and the chance of errors. Therefore, the official results and qrels use expanded qrels this year.

\section{Reusability of test collection}
\label{sec:reusability}

\begin{figure}
    \centering
    \includegraphics[width=0.9\linewidth]{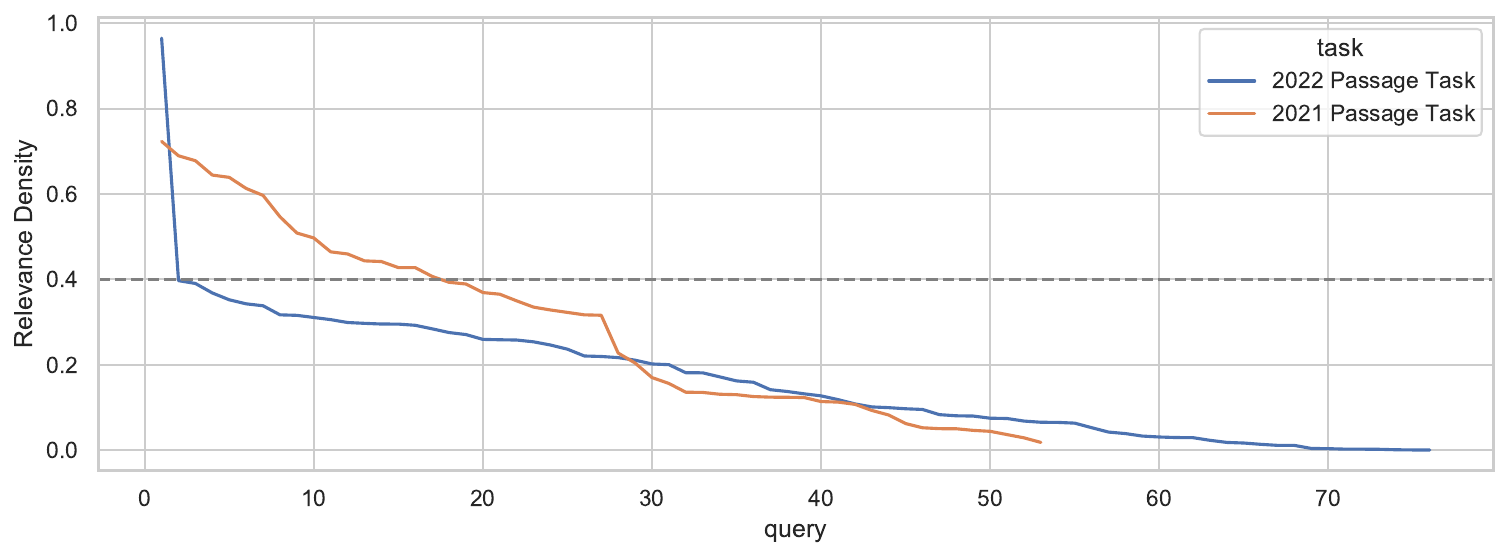}
    \caption{Relevance density.}
    \label{fig:relevance-density}
\end{figure}

In the 2021 track, we identified two problems related to having too many relevant results in the corpus \cite{voorhees2022too}. One problem is reusability. Reusing a test collection means using the same corpus, queries and judgments to evaluate a new IR system, but without doing any additional judging. Because there were too many relevant results in 2021 we ran out of judging resources, and there were too many unjudged relevant results. That means when we evaluate the NDCG@10 of the new IR system, and it retrieves some unjudged documents, there is too much uncertainty about whether the unjudged documents are relevant. 

Our preferred way to handle this is to judge the pools and iteratively judge more candidates identified via classifier, until it is becoming hard to find additional relevant results, making us confident that we can evaluate a new system by correctly assuming that unjudged results are irrelevant. Based on simulation of leaving out certain runs, we established a rule of thumb for reaching sufficiently complete judgments, that the a query's relevance density should be 0.4 or lower. The relevance density is the proportion of judgments that are positive, using the binarization scheme described in Section~\ref{sec:task}.

The other problem with having too many relevant is the saturation of IR metrics. For example, if most IR systems are already getting a Precision@10 of 1.0 for a query, then the query is not useful in evaluation, particularly to identify small but significant differences in relevance between top-performing systems.

To handle these problems of reusability and saturation, we took a number of steps in 2022. Rather than doing separate judging efforts for passage and document tasks, we put all our judging resources into the passage task, and inferred document labels based on passage labels. This made it less likely that we would run out of judging resources before reaching our 0.4 relevance density threshold. We also ran a deduping algorithm, to avoid wasting judging resources on judging the same passage multiple times. 

We also changed the query selection. The original MS MARCO data had one million queries, for which we had Bing's top-10 passages. The passage retrieval was of quite high quality and the passages were deduped by Bing. In all versions of MS MARCO, these Bing results were used to populate the corpus, even in v2 where we included as many URLs from Bing as were available. This also allowed us to evaluate using MS MARCO sparse labels because our test queries had at least one positively labeled result, and every v1 and v2 corpus was constructed to include as many Bing results as possible.

The change in query selection was to use some queries that were run through the same MS MARCO annotation, but the Bing top-10 was not used in corpus construction in v1 or v2. This means we no longer can evaluate the MS MARCO reciprocal rank, because although we have sparse labels, they are not in the corpus. It does make the queries more difficult, because rather than having 10 non-duplicated passages and their source URLs in the corpus, we now are not guaranteed to have any such results. This also makes the task more realistic, because in a real-world IR task we are not guaranteed that the corpus was augmented with Bing results, of course. 




Our first analysis is of the relevance density, to see how many queries are now below our 0.4 rule of thumb indicating ``sufficiently complete'' judgments. Figure~\ref{fig:relevance-density} compares the passage task in 2022 to 2021. Sorting the queries each year by their relevance density, we can see that 2022 had more queries judged in total, and none of them had relevance density of greater than 0.4. They all met the stopping condition. By contrast in 2021 there were 17 topics that didn't reach the stopping condition. For more detail on the reusability of 2021 data, focused on the 2021 document judging, see \cite{voorhees2022too}.


This year there were 24,004 passage judgments, and although it could have theoretically been possible for none of those judgments to be on duplicate passages, around 2\% of judgments this year were duplicates. Last year there were 10,828 judgments, 1715 of which were duplicates, around 15\%. So, if we hadn't done deduping we could have had 64 topics rather than 76. More analysis is needed to understand how much extra statistical power we got, avoiding spending judging resources on duplicates, but the task would still have been possible without deduping. By contrast, had we kept the document task and spent half our budget on it, we could have had 38 topics reach their stopping condition. This is not enough, since we normally hope to have at least 50 topics for each task. We may have included some topics that didn't reach the stopping condition, having a relevance density plot more like the 2021 curve in Figure~\ref{fig:relevance-density}. So deduping was helpful, but eliminating document judging was crucial.


\begin{figure}
    \centering
\includegraphics[width=0.7\linewidth]{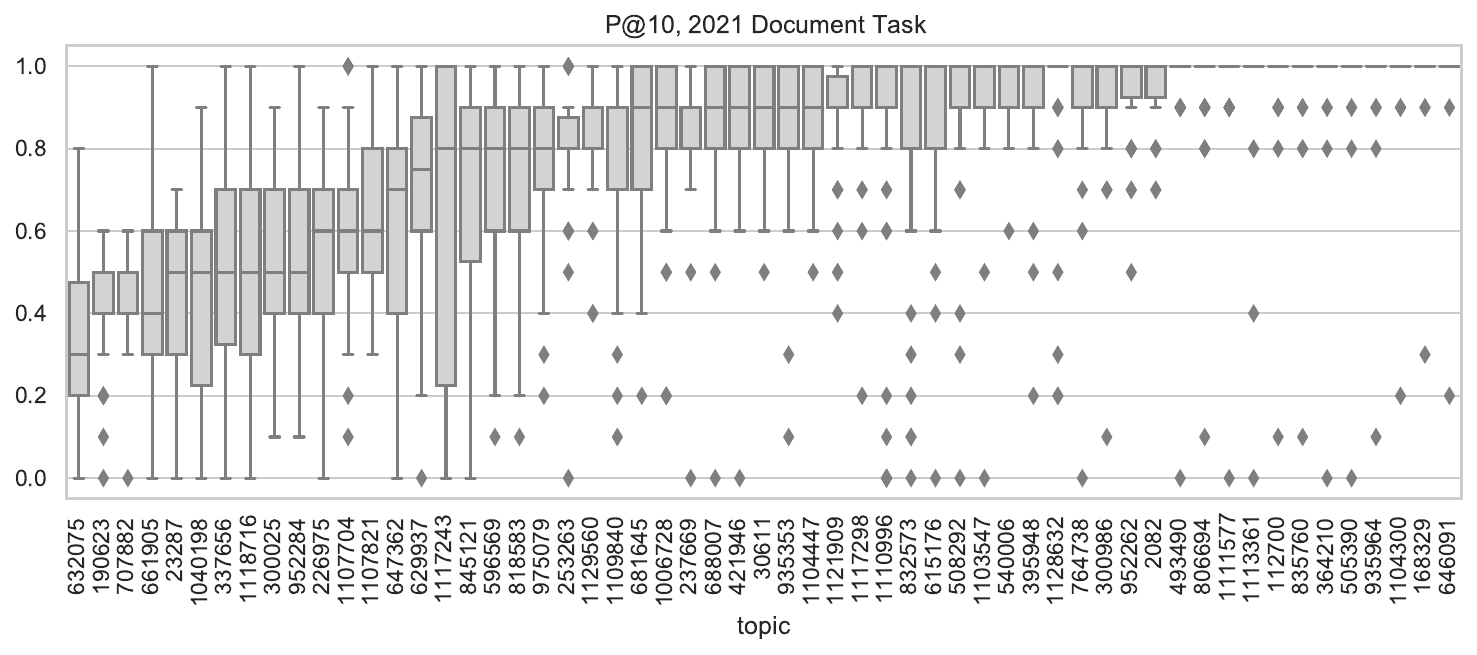}
\includegraphics[width=0.7\linewidth]{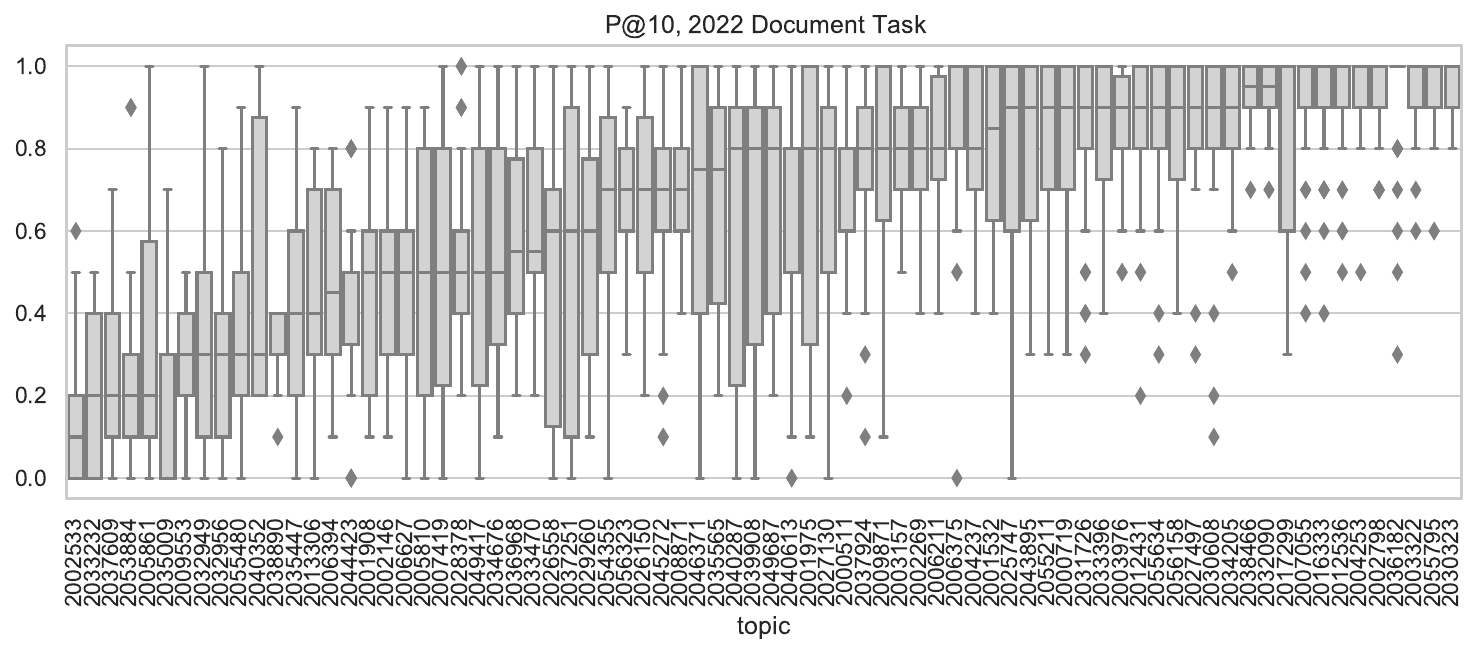}
\includegraphics[width=0.7\linewidth]{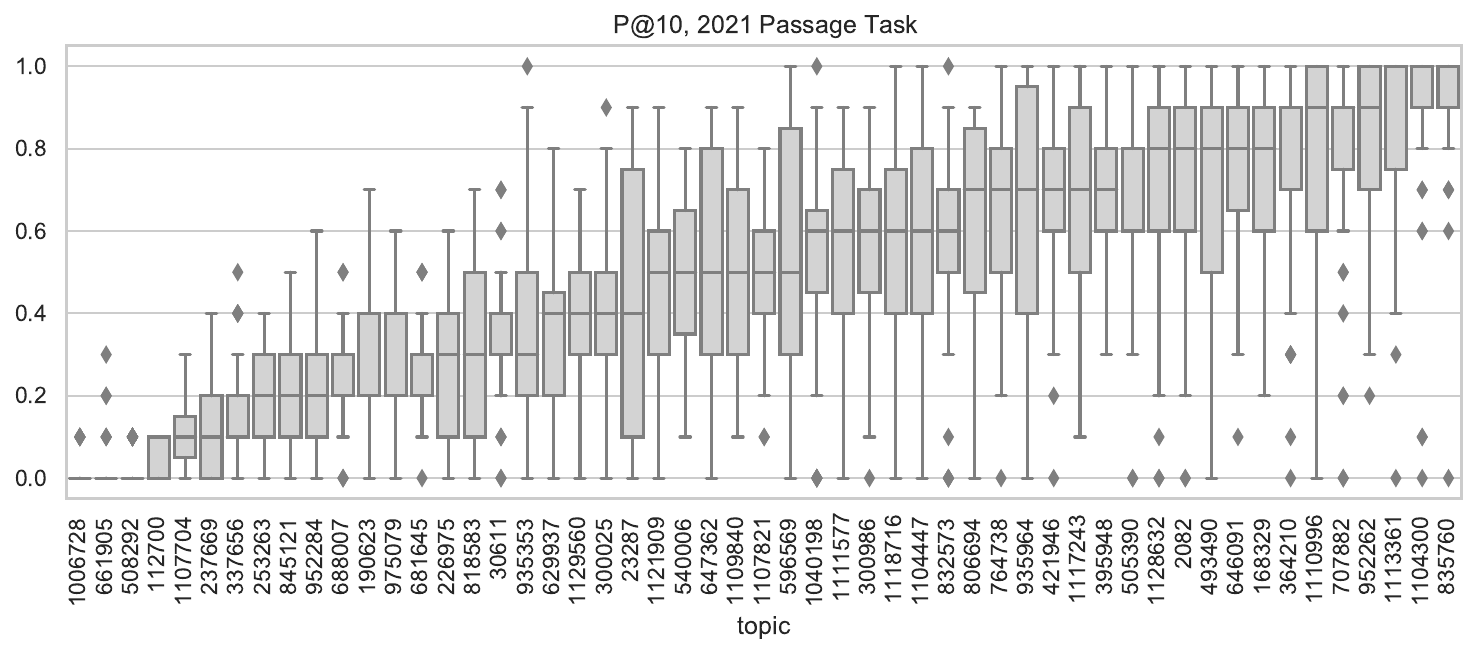}
\includegraphics[width=0.7\linewidth]{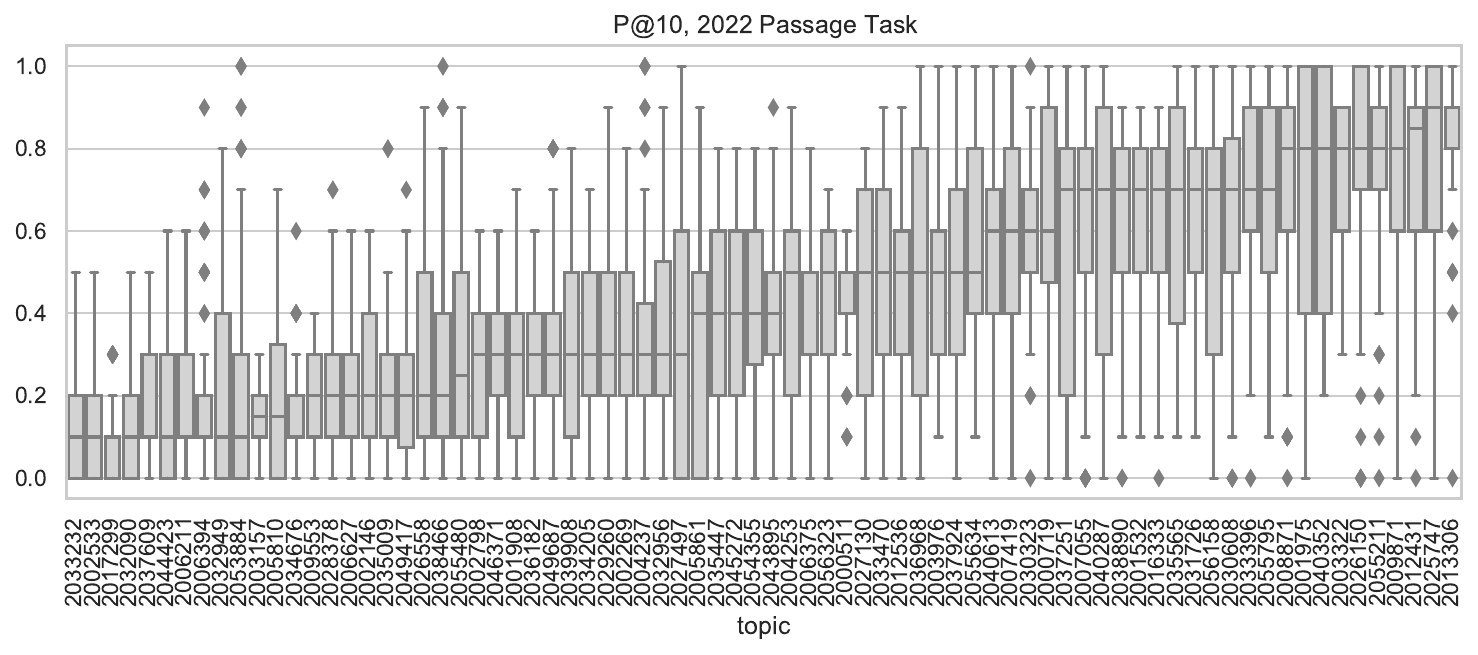}
    \caption{Precision@10 distribution per query.}
    \label{fig:saturation-P_10}
\end{figure}

\begin{figure}
    \centering
\includegraphics[width=0.7\linewidth]{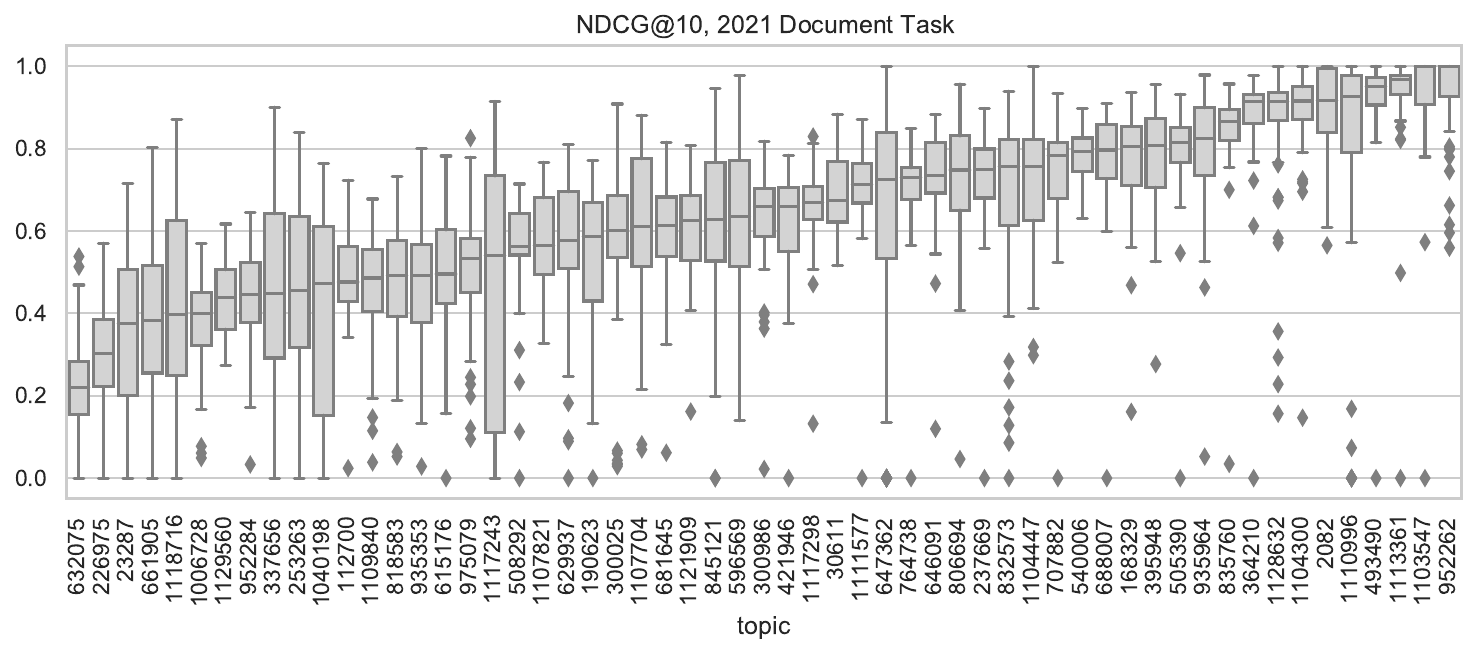}
\includegraphics[width=0.7\linewidth]{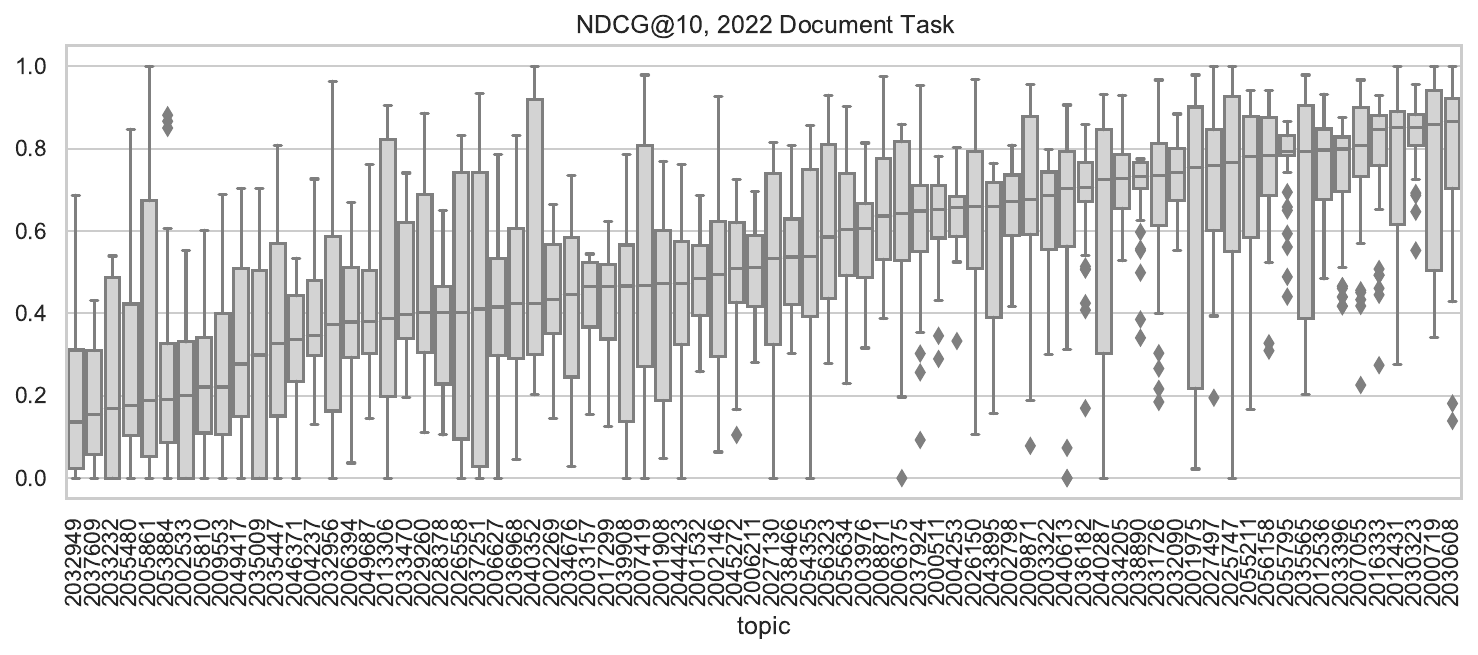}
\includegraphics[width=0.7\linewidth]{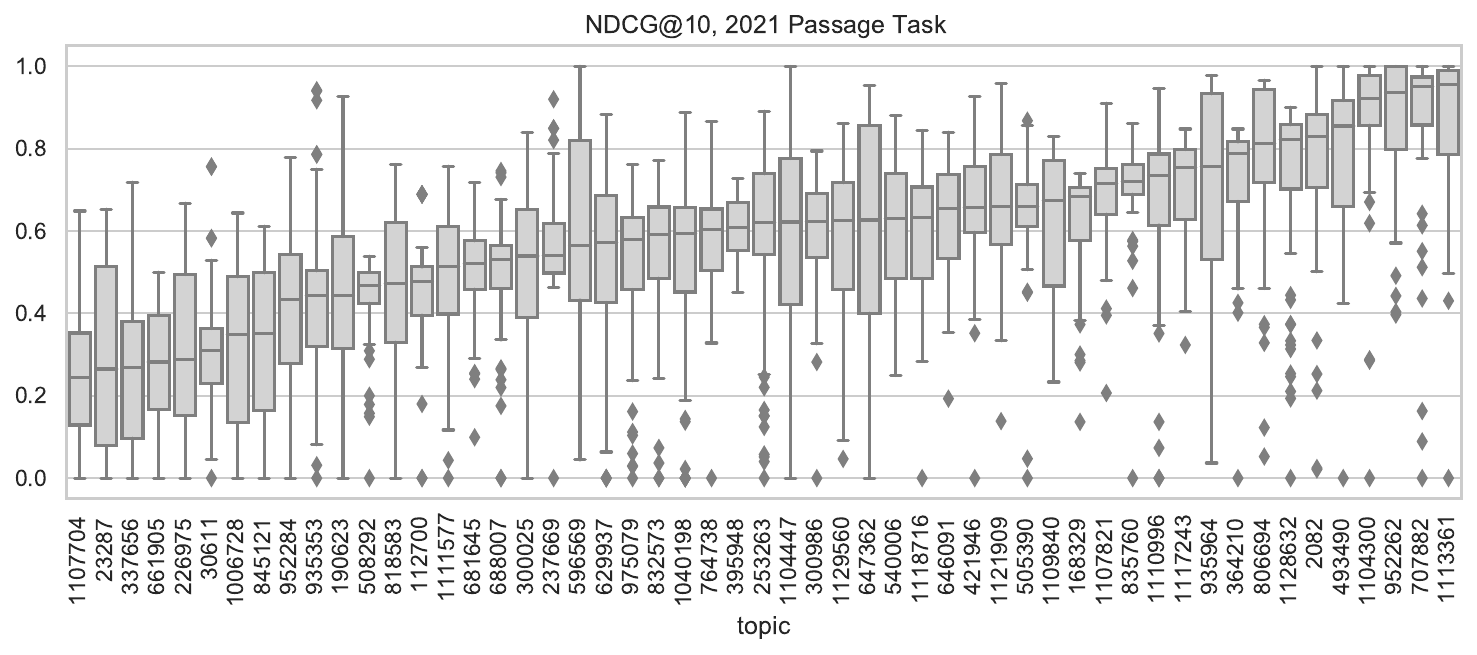}
\includegraphics[width=0.7\linewidth]{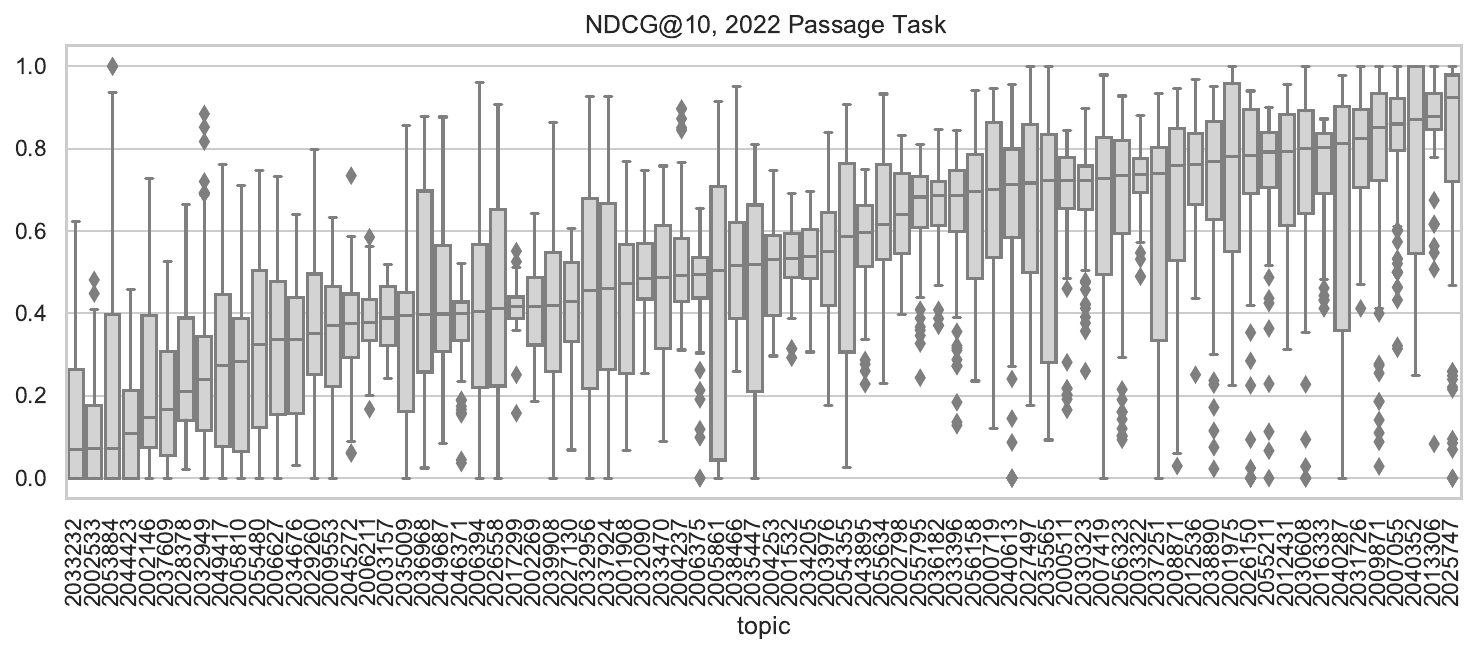}
    \caption{NDCG@10 distribution per query.}
    \label{fig:saturation-ndcg_cut_10}
\end{figure}

To understand the saturation due to too many relevant, consider the per-query metrics of all runs. Considering Precision@10 (Figure~\ref{fig:saturation-P_10}) we replicate the figure from \cite{voorhees2022too} showing that the 2021 document task has several queries where the median is 1.0. In the 2022 task, there are still some queries with this property, but fewer of them. In general there are more queries where the top runs have different Precision@10. We note though that the document task used different judging in the two years, with direct judging of documents in 2021, and inferred document juding in 2022.

For the passage task, which used the same judging scheme both years, we also see a few queries in 2021 that had a median P@10 of 1.0. In 2022 no query has median 1.0. Figure~\ref{fig:saturation-ndcg_cut_10} shows the same analysis for NDCG@10.

\section{Conclusion}
\label{sec:conclusion}
This is the fourth year of the TREC Deep Learning track.
This year the goal was to create a complete collection which could reliably be used to evaluate the performance of different retrieval methods in the passage ranking task. By leveraging a set of harder topics, focused judgement, and passage deduplication the 2022 passage collection is a reusable collection. 
We also continued to observe healthy participation in the track although the number of participating groups reduced slightly this year due to the delay in releasing the test queries.
Deep learning models with large scale pretraining continued to outperform traditional retrieval methods, and single stage retrieval with deep models seems to gain some more ground this year.
This report summarizes our analysis of submitted runs and the observed (mostly positive) impact of the changes in the track this year on building a more complete and consequently more reusable test collections.

\bibliographystyle{plainnat}
\bibliography{bibtex}

\newpage
\appendix

\section{Results Including Baselines}
\label{sec:appendix}

Baseline runs are included to enrich the pools and increase the diversity of approaches used in the evaluation. Baselines are not included in the main results tables, but they are included in our other analysis and in the tables in this appendix.

\begin{table}[h]
\caption{Summary of results for passage ranking runs. Including baseline runs. (1/2)}
\scriptsize
\centering
\begin{tabular}{lllllllrlr}
\toprule
run &      group &   subtask & neural &   stage & dense ret. & baseline &  NDCG@10 & NCG@100 &      AP \\
\midrule
pass3                   &        Ali &  fullrank &   nnlm &   multi &        yes &       no &   0.7184 &  0.4313 &  0.2818 \\
pass2                   &        Ali &  fullrank &   nnlm &   multi &        yes &       no &   0.7105 &  0.4007 &  0.2577 \\
pass1                   &        Ali &  fullrank &   nnlm &   multi &        yes &       no &   0.7050 &  0.4007 &  0.2442 \\
\midrule
cip\_f2\_r                &        CIP &  fullrank &   nnlm &   multi &        yes &       no &   0.5860 &  0.3393 &  0.1761 \\
cip\_f3\_r                &        CIP &  fullrank &   nnlm &   multi &        yes &       no &   0.5852 &  0.3266 &  0.1708 \\
cip\_f1                  &        CIP &  fullrank &   nnlm &  single &        yes &       no &   0.5121 &  0.3393 &  0.1469 \\
cip\_f1\_r                &        CIP &  fullrank &   nnlm &   multi &        yes &       no &   0.5072 &  0.3563 &  0.1622 \\
cip\_f2                  &        CIP &  fullrank &   nnlm &  single &        yes &       no &   0.4997 &  0.3563 &  0.1429 \\
cip\_r2                  &        CIP &    rerank &   nnlm &   multi &         no &       no &   0.4975 &  0.1826 &  0.0891 \\
cip\_f3                  &        CIP &  fullrank &   nnlm &  single &        yes &       no &   0.4840 &  0.3266 &  0.1357 \\
cip\_r3                  &        CIP &    rerank &   nnlm &   multi &         no &       no &   0.4669 &  0.1826 &  0.0795 \\
cip\_r1                  &        CIP &    rerank &   nnlm &   multi &         no &       no &   0.4320 &  0.1826 &  0.0719 \\
\midrule
tuvienna-pas-col        &    DOSSIER &  fullrank &   nnlm &  single &        yes &       no &   0.5386 &  0.3331 &  0.1677 \\
tuvienna-pas-unicol     &    DOSSIER &  fullrank &   nnlm &  single &        yes &       no &   0.5231 &  0.3212 &  0.1518 \\
\midrule
Infosense-2             &  InfoSense &    rerank &   nnlm &   multi &         no &       no &   0.4848 &  0.1826 &  0.0846 \\
Infosense-1             &  InfoSense &    rerank &   nnlm &   multi &         no &       no &   0.4832 &  0.1826 &  0.0830 \\
\midrule
NLE\_SPLADE\_CBERT\_DT5\_RR &        NLE &  fullrank &   nnlm &   multi &         no &       no &   0.7145 &  0.4592 &  0.2950 \\
NLE\_SPLADE\_CBERT\_RR     &        NLE &  fullrank &   nnlm &   multi &         no &       no &   0.7141 &  0.4565 &  0.2963 \\
NLE\_SPLADE\_RR           &        NLE &  fullrank &   nnlm &   multi &         no &       no &   0.7092 &  0.4589 &  0.2977 \\
SPLADE\_ENSEMBLE\_PP\_RCIO &        NLE &  fullrank &   nnlm &   multi &         no &      yes &   0.5991 &  0.3823 &  0.2005 \\
SPLADE\_PP\_ED\_RCIO       &        NLE &  fullrank &   nnlm &   multi &         no &      yes &   0.5917 &  0.3748 &  0.1923 \\
SPLADE\_PP\_SD\_RCIO       &        NLE &  fullrank &   nnlm &   multi &         no &      yes &   0.5897 &  0.3748 &  0.1968 \\
SPLADE\_ENSEMBLE\_PP      &        NLE &  fullrank &   nnlm &   multi &         no &      yes &   0.5789 &  0.3784 &  0.1862 \\
SPLADE\_PP\_ED            &        NLE &  fullrank &   nnlm &   multi &         no &      yes &   0.5786 &  0.3680 &  0.1801 \\
SPLADE\_PP\_SD            &        NLE &  fullrank &   nnlm &   multi &         no &      yes &   0.5705 &  0.3702 &  0.1846 \\
SPLADE\_EFF\_V            &        NLE &  fullrank &   nnlm &   multi &         no &      yes &   0.5509 &  0.3419 &  0.1631 \\
SPLADE\_EFF\_V\_RCIO       &        NLE &  fullrank &   nnlm &   multi &         no &      yes &   0.5452 &  0.3362 &  0.1725 \\
NLE\_ENSEMBLE\_SUM        &        NLE &    rerank &   nnlm &   multi &         no &       no &   0.5286 &  0.1826 &  0.0948 \\
NLE\_ENSEMBLE\_CONDORCET  &        NLE &    rerank &   nnlm &   multi &         no &       no &   0.5284 &  0.1826 &  0.0943 \\
SPLADE\_EFF\_VI-BT        &        NLE &  fullrank &   nnlm &   multi &         no &      yes &   0.5271 &  0.3210 &  0.1452 \\
NLE\_T0pp                &        NLE &    rerank &   nnlm &   multi &         no &       no &   0.5102 &  0.1826 &  0.0881 \\
SPLADE\_EFF\_VI-BT\_RCIO   &        NLE &  fullrank &   nnlm &   multi &         no &      yes &   0.5084 &  0.3061 &  0.1452 \\
\midrule
2systems                &        UGA &  fullrank &   nnlm &   multi &        yes &       no &   0.5991 &  0.2958 &  0.1622 \\
unicoil\_reranked        &        UGA &  fullrank &   nnlm &   multi &        yes &       no &   0.5910 &  0.2958 &  0.1605 \\
6systems                &        UGA &  fullrank &   nnlm &   multi &        yes &       no &   0.5783 &  0.3218 &  0.1604 \\
4systems                &        UGA &  fullrank &   nnlm &   multi &        yes &       no &   0.5761 &  0.2959 &  0.1530 \\
c47                     &        UGA &  fullrank &   nnlm &   multi &        yes &       no &   0.5701 &  0.2958 &  0.1493 \\
hierarchcal\_combination &        UGA &  fullrank &   nnlm &   multi &        yes &       no &   0.5696 &  0.3554 &  0.1655 \\
graph\_colbert           &        UGA &  fullrank &   nnlm &   multi &        yes &       no &   0.5482 &  0.3545 &  0.1656 \\
fused\_3runs             &        UGA &    rerank &   nnlm &   multi &        yes &       no &   0.5094 &  0.1826 &  0.0901 \\
fused\_2runs             &        UGA &    rerank &   nnlm &   multi &        yes &       no &   0.5060 &  0.1826 &  0.0895 \\
hierarchical\_2runs      &        UGA &    rerank &   nnlm &   multi &        yes &       no &   0.5001 &  0.1826 &  0.0885 \\
\midrule
uogtr\_se                &      UoGTr &  fullrank &   nnlm &   multi &         no &      yes &   0.6510 &  0.3826 &  0.2252 \\
uogtr\_se\_gb             &      UoGTr &  fullrank &   nnlm &   multi &         no &       no &   0.6508 &  0.3825 &  0.2252 \\
uogtr\_se\_gt             &      UoGTr &  fullrank &   nnlm &   multi &         no &       no &   0.6508 &  0.3824 &  0.2256 \\
uogtr\_e\_gb              &      UoGTr &  fullrank &   nnlm &   multi &        yes &       no &   0.6501 &  0.3818 &  0.2257 \\
uogtr\_be\_gb             &      UoGTr &  fullrank &   nnlm &   multi &         no &       no &   0.6480 &  0.3558 &  0.2113 \\
uogtr\_be                &      UoGTr &  fullrank &   nnlm &   multi &         no &      yes &   0.6235 &  0.3252 &  0.1896 \\
uogtr\_e\_cprf\_t5         &      UoGTr &  fullrank &   nnlm &   multi &        yes &       no &   0.6182 &  0.3621 &  0.2061 \\
uogtr\_s                 &      UoGTr &  fullrank &   nnlm &   multi &         no &      yes &   0.5697 &  0.3699 &  0.1831 \\
uogtr\_s\_cprf            &      UoGTr &  fullrank &   nnlm &   multi &        yes &       no &   0.5682 &  0.3501 &  0.1866 \\
uogtr\_c                 &      UoGTr &  fullrank &   nnlm &  single &        yes &      yes &   0.5217 &  0.2419 &  0.1319 \\
uogtr\_t\_cprf            &      UoGTr &  fullrank &   nnlm &   multi &        yes &       no &   0.5078 &  0.3250 &  0.1646 \\
uogtr\_c\_cprf            &      UoGTr &  fullrank &   nnlm &   multi &        yes &       no &   0.5075 &  0.2488 &  0.1355 \\
uogtr\_dph\_bo1           &      UoGTr &  fullrank &   trad &   multi &         no &      yes &   0.3050 &  0.1836 &  0.0433 \\
uogtr\_dph               &      UoGTr &  fullrank &   trad &  single &         no &      yes &   0.2905 &  0.1754 &  0.0410 \\
\bottomrule
\end{tabular}

\label{tab:passage_ranking_all1}
\end{table}

\begin{table}[]
\caption{Summary of results for passage ranking runs. Including baseline runs. (2/2)}
\scriptsize
\centering
\begin{tabular}{lllllllrlr}
\toprule
run &      group &   subtask & neural &   stage & dense ret. & baseline &  NDCG@10 & NCG@100 &      AP \\
\midrule
webis-dl-duot5-g        &      Webis &  fullrank &   nnlm &   multi &         no &       no &   0.5314 &  0.1501 &  0.0887 \\
webis-dl-duot5          &      Webis &  fullrank &   nnlm &   multi &         no &       no &   0.4972 &  0.1501 &  0.0800 \\
webis-dl-duot5-aug-1    &      Webis &  fullrank &   nnlm &   multi &         no &       no &   0.4925 &  0.1226 &  0.0781 \\
webis-dl-duot5-aug-2    &      Webis &  fullrank &   nnlm &   multi &         no &       no &   0.4885 &  0.1226 &  0.0759 \\
\midrule
f\_sum\_mdt5              &     h2oloo &  fullrank &   nnlm &   multi &        yes &       no &   0.7030 &  0.3993 &  0.2698 \\
p\_d2q\_bm25rocchio\_mdt5  &     h2oloo &  fullrank &   nnlm &   multi &         no &      yes &   0.6933 &  0.3724 &  0.2374 \\
p\_d2q\_bm25rocchio\_mt5   &     h2oloo &  fullrank &   nnlm &   multi &         no &      yes &   0.6282 &  0.3724 &  0.2122 \\
p\_dhr                   &     h2oloo &  fullrank &   nnlm &  single &        yes &       no &   0.5524 &  0.3420 &  0.1662 \\
p\_tct                   &     h2oloo &  fullrank &   nnlm &  single &        yes &      yes &   0.5329 &  0.3155 &  0.1540 \\
p\_agg                   &     h2oloo &  fullrank &   nnlm &  single &        yes &       no &   0.5282 &  0.3119 &  0.1461 \\
p\_unicoil\_exp\_rocchio   &     h2oloo &  fullrank &   nnlm &  single &         no &      yes &   0.4886 &  0.3069 &  0.1225 \\
p\_unicoil\_exp           &     h2oloo &  fullrank &   nnlm &  single &         no &      yes &   0.4614 &  0.2957 &  0.1050 \\
p\_unicoil\_noexp\_rocchio &     h2oloo &  fullrank &   nnlm &  single &         no &      yes &   0.4164 &  0.2552 &  0.0974 \\
p\_unicoil\_noexp         &     h2oloo &  fullrank &   nnlm &  single &         no &      yes &   0.4077 &  0.2383 &  0.0754 \\
paug\_d2q\_bm25rocchio    &     h2oloo &  fullrank &   nnlm &  single &         no &      yes &   0.3801 &  0.2527 &  0.0860 \\
paug\_d2q\_bm25rm3        &     h2oloo &  fullrank &   nnlm &  single &         no &      yes &   0.3754 &  0.2478 &  0.0818 \\
p\_d2q\_bm25rocchio       &     h2oloo &  fullrank &   nnlm &  single &         no &      yes &   0.3712 &  0.2698 &  0.0868 \\
p\_d2q\_bm25rm3           &     h2oloo &  fullrank &   nnlm &  single &         no &      yes &   0.3704 &  0.2695 &  0.0864 \\
paug\_d2q\_bm25           &     h2oloo &  fullrank &   nnlm &  single &         no &      yes &   0.3609 &  0.2520 &  0.0735 \\
p\_d2q\_bm25              &     h2oloo &  fullrank &   nnlm &  single &         no &      yes &   0.3599 &  0.2535 &  0.0748 \\
paug\_bm25               &     h2oloo &  fullrank &   trad &  single &         no &      yes &   0.2742 &  0.1684 &  0.0346 \\
p\_bm25rocchio           &     h2oloo &  fullrank &   trad &  single &         no &      yes &   0.2741 &  0.1820 &  0.0340 \\
p\_bm25rm3               &     h2oloo &  fullrank &   trad &  single &         no &      yes &   0.2724 &  0.1732 &  0.0326 \\
p\_bm25                  &     h2oloo &  fullrank &   trad &  single &         no &      yes &   0.2692 &  0.1826 &  0.0325 \\
paug\_bm25rocchio        &     h2oloo &  fullrank &   trad &  single &         no &      yes &   0.2593 &  0.1506 &  0.0314 \\
paug\_bm25rm3            &     h2oloo &  fullrank &   trad &  single &         no &      yes &   0.2591 &  0.1520 &  0.0318 \\
\midrule
IELab-3MP-UT            &      ielab &  fullrank &   nnlm &  single &         no &       no &   0.4658 &  0.2888 &  0.1101 \\
IELab-3MP-RBC           &      ielab &  fullrank &   nnlm &  single &         no &       no &   0.4368 &  0.3220 &  0.1013 \\
IELab-3MP-DI            &      ielab &  fullrank &   nnlm &  single &         no &       no &   0.4148 &  0.2663 &  0.0832 \\
IELab-3MP-DT5           &      ielab &  fullrank &   nnlm &  single &         no &      yes &   0.3620 &  0.2480 &  0.0713 \\
\midrule
srchvrs\_pz2\_colb2       &    srchvrs &  fullrank &   nnlm &   multi &        yes &       no &   0.6630 &  0.3660 &  0.2160 \\
srchvrs\_ptn1\_colb2      &    srchvrs &  fullrank &   nnlm &   multi &        yes &       no &   0.6562 &  0.3660 &  0.2066 \\
srchvrs\_ptn2\_colb2      &    srchvrs &  fullrank &   nnlm &   multi &        yes &       no &   0.6448 &  0.3660 &  0.2002 \\
srchvrs\_pz1\_colb2       &    srchvrs &  fullrank &   nnlm &   multi &         no &       no &   0.6414 &  0.3501 &  0.2096 \\
srchvrs\_ptn1\_lcn\_colb2  &    srchvrs &  fullrank &   nnlm &   multi &         no &       no &   0.6367 &  0.3501 &  0.1996 \\
srchvrs\_p2\_colb2        &    srchvrs &  fullrank &   nnlm &   multi &        yes &       no &   0.6010 &  0.3492 &  0.1745 \\
srchvrs\_p1\_colb2        &    srchvrs &  fullrank &   nnlm &   multi &         no &       no &   0.5818 &  0.3400 &  0.1723 \\
srchvrs\_ptn3\_colb2      &    srchvrs &  fullrank &   nnlm &   multi &        yes &       no &   0.5800 &  0.3660 &  0.1687 \\
srchvrs\_p\_bm25\_mdl1     &    srchvrs &  fullrank &   trad &   multi &         no &      yes &   0.3194 &  0.2040 &  0.0414 \\
srchvrs\_p\_bm25f\_mdl1    &    srchvrs &  fullrank &   trad &   multi &         no &      yes &   0.3161 &  0.2080 &  0.0415 \\
srchvrs\_p\_bm25f         &    srchvrs &  fullrank &   trad &   multi &         no &      yes &   0.3153 &  0.2029 &  0.0405 \\
srchvrs\_p\_bm25          &    srchvrs &  fullrank &   trad &  single &         no &      yes &   0.2911 &  0.1801 &  0.0340 \\
\midrule
yorku22a                &    yorku22 &  fullrank &   nnlm &   multi &        yes &       no &   0.6089 &  0.3747 &  0.2003 \\
yorku22b                &    yorku22 &  fullrank &   nnlm &  single &         no &       no &   0.5076 &  0.2692 &  0.1130 \\
\bottomrule
\end{tabular}

\label{tab:passage_ranking_all2}
\end{table}

\begin{table}[]
\caption{Summary of results for document ranking runs. Including baseline runs.}
\scriptsize
\centering
\begin{tabular}{lllllllrlr}
\toprule
run &          group &   subtask & neural &   stage & dense ret. & baseline &  NDCG@10 & NCG@100 &      AP \\
\midrule
doc3                      &            Ali &  fullrank &   nnlm &   multi &        yes &       no &   0.7488 &  0.5246 &  0.2997 \\
doc1                      &            Ali &  fullrank &   nnlm &   multi &        yes &       no &   0.4936 &  0.4739 &  0.2154 \\
doc2                      &            Ali &  fullrank &   nnlm &   multi &        yes &       no &   0.4589 &  0.4739 &  0.2030 \\
ceqe\_custom\_rerank        &  CERTH\_ITI\_M4D &  fullrank &   nnlm &   multi &        yes &       no &   0.3811 &  0.2599 &  0.1090 \\
rm3\_term\_filter\_rerank    &  CERTH\_ITI\_M4D &  fullrank &   nnlm &   multi &        yes &       no &   0.3611 &  0.2425 &  0.1049 \\
\midrule
tuvienna                  &        DOSSIER &  fullrank &   nnlm &  single &        yes &       no &   0.4868 &  0.3043 &  0.1294 \\
tuvienna-unicol           &        DOSSIER &  fullrank &   nnlm &  single &        yes &       no &   0.4830 &  0.2985 &  0.1232 \\
\midrule
NLE\_SPLADE\_RR\_D           &            NLE &  fullrank &   nnlm &   multi &         no &       no &   0.7611 &  0.5787 &  0.3453 \\
NLE\_SPLADE\_CBERT\_RR\_D     &            NLE &  fullrank &   nnlm &   multi &         no &       no &   0.7601 &  0.5716 &  0.3387 \\
NLE\_SPLADE\_CBERT\_DT5\_RR\_D &            NLE &  fullrank &   nnlm &   multi &         no &       no &   0.7598 &  0.5782 &  0.3405 \\
SPLADE\_ENSEMBLE\_PP\_RCIO\_D &            NLE &  fullrank &   nnlm &   multi &         no &      yes &   0.6584 &  0.5216 &  0.2933 \\
SPLADE\_PP\_ED\_RCIO\_D       &            NLE &  fullrank &   nnlm &   multi &         no &      yes &   0.6566 &  0.5122 &  0.2864 \\
SPLADE\_PP\_SD\_RCIO\_D       &            NLE &  fullrank &   nnlm &   multi &         no &      yes &   0.6471 &  0.5157 &  0.2835 \\
SPLADE\_PP\_ED\_D            &            NLE &  fullrank &   nnlm &   multi &         no &      yes &   0.6448 &  0.4951 &  0.2656 \\
SPLADE\_ENSEMBLE\_PP\_D      &            NLE &  fullrank &   nnlm &   multi &         no &      yes &   0.6402 &  0.5090 &  0.2740 \\
SPLADE\_PP\_SD\_D            &            NLE &  fullrank &   nnlm &   multi &         no &      yes &   0.6269 &  0.5035 &  0.2692 \\
SPLADE\_EFF\_V\_D            &            NLE &  fullrank &   nnlm &   multi &         no &      yes &   0.6094 &  0.4550 &  0.2345 \\
SPLADE\_EFF\_V\_RCIO\_D       &            NLE &  fullrank &   nnlm &   multi &         no &      yes &   0.6031 &  0.4780 &  0.2524 \\
NLE\_ENSEMBLE\_SUM\_doc      &            NLE &  fullrank &   nnlm &   multi &         no &       no &   0.5918 &  0.2593 &  0.1619 \\
NLE\_ENSEMBLE\_CONDORCE\_doc &            NLE &  fullrank &   nnlm &   multi &         no &       no &   0.5882 &  0.2593 &  0.1609 \\
NLE\_T0pp\_doc              &            NLE &  fullrank &   nnlm &   multi &         no &       no &   0.5843 &  0.2593 &  0.1587 \\
SPLADE\_EFF\_VI-BT\_D        &            NLE &  fullrank &   nnlm &   multi &         no &      yes &   0.5758 &  0.4333 &  0.2152 \\
SPLADE\_EFF\_VI-BT\_RCIO\_D   &            NLE &  fullrank &   nnlm &   multi &         no &      yes &   0.5755 &  0.4421 &  0.2245 \\
\midrule
plm\_128                   &     UAmsterdam &    rerank &   nnlm &   multi &         no &       no &   0.3387 &  0.2236 &  0.0905 \\
plm\_64                    &     UAmsterdam &    rerank &   nnlm &   multi &         no &       no &   0.3227 &  0.2236 &  0.0909 \\
plm\_512                   &     UAmsterdam &    rerank &   nnlm &   multi &         no &       no &   0.2721 &  0.2236 &  0.0816 \\
\midrule
uogtr\_doc\_dph\_bo1         &          UoGTr &  fullrank &   trad &   multi &         no &      yes &   0.3625 &  0.2921 &  0.1199 \\
uogtr\_doc\_dph             &          UoGTr &  fullrank &   trad &  single &         no &      yes &   0.3603 &  0.2847 &  0.1099 \\
\midrule
srchvrs\_dtn1              &        srchvrs &  fullrank &   nnlm &   multi &        yes &       no &   0.5970 &  0.3492 &  0.1816 \\
srchvrs\_dtn2              &        srchvrs &  fullrank &   nnlm &   multi &        yes &       no &   0.5888 &  0.3492 &  0.1798 \\
srchvrs\_d\_lb2             &        srchvrs &  fullrank &   nnlm &   multi &        yes &       no &   0.5760 &  0.3492 &  0.1777 \\
srchvrs\_d\_lb1             &        srchvrs &  fullrank &   nnlm &   multi &        yes &       no &   0.5754 &  0.3492 &  0.1782 \\
srchvrs\_d\_prd3            &        srchvrs &  fullrank &   nnlm &   multi &        yes &       no &   0.5620 &  0.3492 &  0.1742 \\
srchvrs\_d\_prd1            &        srchvrs &  fullrank &   nnlm &   multi &        yes &       no &   0.5546 &  0.3492 &  0.1705 \\
srchvrs\_d\_lb3             &        srchvrs &  fullrank &   nnlm &   multi &         no &       no &   0.5302 &  0.2748 &  0.1407 \\
srchvrs\_d\_bm25\_pass\_mf    &        srchvrs &  fullrank &   trad &   multi &         no &      yes &   0.4318 &  0.3140 &  0.1340 \\
srchvrs\_d\_bm25\_pass\_mdl1  &        srchvrs &  fullrank &   trad &   multi &         no &      yes &   0.4269 &  0.3192 &  0.1336 \\
srchvrs\_d\_bm25\_p\_mf\_mdl1  &        srchvrs &  fullrank &   trad &   multi &         no &      yes &   0.4243 &  0.3035 &  0.1286 \\
srchvrs\_d\_bm25\_mf\_mdl1    &        srchvrs &  fullrank &   trad &   multi &         no &      yes &   0.3883 &  0.2804 &  0.1082 \\
srchvrs\_d\_bm25\_mf         &        srchvrs &  fullrank &   trad &   multi &         no &      yes &   0.3841 &  0.2829 &  0.1116 \\
srchvrs\_d\_bm25\_mdl1       &        srchvrs &  fullrank &   trad &   multi &         no &      yes &   0.3817 &  0.2956 &  0.1157 \\
srchvrs\_d\_bm25            &        srchvrs &  fullrank &   trad &  single &         no &      yes &   0.3388 &  0.2742 &  0.1048 \\
\bottomrule
\end{tabular}

\label{tab:document_ranking_all}
\end{table}

\end{document}